\documentclass[11pt,a4paper]{article}
\pdfoutput=1
\usepackage{jheppub}
\usepackage{graphics}      
\usepackage{graphicx}      
\usepackage[font=small]{caption}
\usepackage{subcaption}
\usepackage{longtable}     
\usepackage{url}           
\usepackage{bm}      
\graphicspath{{figs/}}

\title{Entanglement Entropy in Jammed CFTs}
\author[a]{Eric Mefford}
\affiliation[a]{Department of Physics\\
University of California, Santa Barbara\\
Santa Barbara, CA 93106, USA}
\emailAdd{mefford@physics.ucsb.edu}
\abstract{We construct solutions to the Einstein equations for asymptotically locally Anti-de Sitter spacetimes with four, five, and six dimensional Reissner-Nordstr{\"o}m boundary metrics. These spacetimes are gravitational duals to ``jammed" CFTs on those backgrounds at infinite N and strong coupling. For these spacetimes, we calculate the boundary stress tensor as well as compute entanglement entropies for ball shaped regions as functions of the boundary black hole temperature $T_{BH}$. From this, we see how the CFT prevents heat flow from the black hole to the vacuum at spatial infinity. We also compute entanglement entropies for a three dimensional boundary black hole using the AdS C-metric. We compare our results to previous work done in similar spacetimes.}

\begin{document}

\date{\today}

\maketitle
\flushbottom

\section{Introduction}
The study of quantum field theories on curved spacetimes has historically been a source of both deep and enigmatic discoveries in theoretical physics. For instance, the analysis of an accelerated observer in Minkowski space showed that the field theory in the observer's frame and the field theory in Minkowski spacetime do not share a common vacuum \cite{Fulling:1972md, Davies:1974th, Unruh:1976db}. Furthermore, theories invariant under metric rescaling (Weyl transformations) have classically traceless stress tensors. However, when these theories are quantized on a curved manifold in even spacetime dimensions, it is found that at one-loop order, the trace picks up contributions proportional to geometric invariants of the spacetime \cite{Capper:1974ic}.  Possibly the most interesting and perplexing discovery, however, is that black holes, when analyzed quantum mechanically, are not ever-growing cosmic sinks but rather radiate away their energy with a nearly thermal spectrum \cite{Hawking:1974rv}. This discovery has led to new insights into thermodynamics \cite{Bekenstein:1973ur} as well as illuminated fundamental issues in quantum mechanics and the conservation of information \cite{Almheiri:2012rt}. It may not be too surprising to learn that these discoveries are related---for instance, in the context of two dimensional CFTs, Hawking radiation is completely determined by the conformal anomaly \cite{Christensen:1977jc}. On the other hand, it should be noted that the majority of analysis has been in the context of free fields. Recently, the impact of interactions on these phenomena have begun to be explored \cite{Hubeny:2009ru, Hubeny:2009kz, Hubeny:2009rc, Marolf:2013ioa}.

One particularly fruitful avenue for addressing these questions is the AdS/CFT correspondence \cite{Maldacena:1997re, Gubser:1998bc, Witten:1998qj}. Here, one is able to study a strongly interacting d-dimensional $U(N)$ conformal field theory on a fixed spacetime background $\mathcal{B}_d$ by considering a d+1 dimensional solution to Einstein's equations with negative cosmological constant. The d+1 gravitational solution, $\mathcal{M}$, is frequently referred to as ``the bulk." The boundary of $\mathcal{M}$ is conformal to the background spacetime on which the conformal field theory lives. In the infinite $N$ limit, the planar graph contributions to expectation values of field theory operators on $\mathcal{B}_d$ may be obtained by solving classical equations of motion for corresponding matter fields in $\mathcal{M}$ \cite{Gubser:1998bc}. It should be noted that gravity is not dynamical on the boundary. In particular, $\mathcal{B}_d$ serves as a classical background for the field theory, with no backreaction taking place.\footnote{One may extend the AdS/CFT correspondence to address dynamical gravity by imposing Neumann-like boundary conditions for the CFT metric \cite{Compere:2008us}.} This limits some of the questions that may be addressed as $G_d$ is now effectively zero. For instance, questions like the black hole information paradox \cite{Hawking:1976ra, Almheiri:2012rt} for which the black hole not only radiates but also evaporates cannot be addressed by considering a boundary black hole.\footnote{At finite N, however, it is expected that AdS/CFT will give valuable insight into this questions when one considers a bulk black hole dual to a thermal field theory.} Nevertheless, we may still think of the black hole as a heat source for the field theory to explore heat transport and use this to characterize unique phases of the interacting field theory.

To analyze properties of Hawking radiation on the CFT, we construct new five, six, and seven dimensional solutions to the Einstein equations for asymptotically locally Anti-de Sitter spacetimes that have Reissner-Nordstr{\"o}m metrics on the boundary. These new solutions build upon \cite{Hubeny:2009ru, Hubeny:2009kz, Hubeny:2009rc, Marolf:2013ioa} in which the authors considered spacetimes with boundaries $\mathcal{B}_d$ which contained a hyperbolic black hole of size $R_{BH}$ at temperature $T_{BH}$. The hyperbolic black hole spacetimes also contained a black hole in the bulk at temperature $T_\infty$. Generally, the bulk horizon is thought to represent the dual of a thermal state in the field theory at the same temperature. However, because of the presence of the boundary black hole, the authors of \cite{Fischetti:2013hja} consider the bulk black hole as governing a thermal plasma at spatial infinity which serves as a heat sink for the CFT. The boundary horizon then serves as a heat source. Our solutions have a Poincar{\'e} horizon in the bulk so that $T_\infty = 0$. 

There have even been spacetimes, as constructed by \cite{Fischetti:2012ps, Fischetti:2012vt}, with only one Killing vector allowing the CFT to be at a third temperature $T_0$ in a ``detuned" phase. In these so called ``flowing funnels", if $T_0\neq T_{BH}$, the authors of \cite{Fischetti:2012ps, Fischetti:2012vt} state the stress tensor will be singular at the horizon. In our solutions below, the stress tensor is finite, and so we consider our solution ``tuned" with only two temperatures, $T_{BH}$ and $T_\infty$. Even in this case, the authors of \cite{Figueras:2011va} suggest that if $T_{BH}\neq T_{\infty}$, $\mathcal{O}(1/N^2)$ effects in the CFT may introduce singularities at the horizon. In this paper, as we are operating in the planar limit of the field theory, we will not be able to definitively distinguish between these two scenarios. However, we will see that field theory observables are markedly different near the horizon than they are far away, and that the near horizon observables have a strong dependence on $T_{BH}$. 

The presence of two temperatures on the boundary allows one to explore different potential phases of Hawking radiation that the authors of \cite{Marolf:2013ioa} suggest correspond to different vacuum states of the CFT. Varying the dimensionless parameter $R_{BH}T_\infty$ corresponds to adjusting the relative distance between the bulk and boundary horizons. Heuristically, we can see this as follows. Because the spacetimes we construct will correspond to asymptotically flat, spherically symmetric boundary spacetimes, we can consider the bulk horizons to be asymptotically planar. In terms of the so-called ``Fefferman-Graham" coordinate \cite{Fefferman:2007rka}, $z$, for which the boundary of our bulk spacetime is at $z=0$, very far from the rotation axis, the bulk horizon location will roughly be at a location $z_h=1/T_\infty$. Furthermore, the maximum $z$ location to which the boundary horizon extends into the bulk is roughly $z_b=R_{BH}$. With this in mind, when $R_{BH}T_\infty\ll1$, we are in a so-called ``droplet phase" in which the bulk and boundary horizons are disconnected and very far separated. As this corresponds to a large $T_{BH}/T_\infty$, it is seen that there is very little heat transport in the CFT, a scenario the authors of \cite{Fischetti:2013hja} refer to as ``jammed." As we take $R_{BH}T_\infty \to 1$, the separation between the boundary and bulk black holes goes to zero. This may lead to a phase transition to a so-called ``funnel phase" in which the bulk and boundary black holes are connected. In this phase, there is only one Killing horizon, and so $T_{BH}$ = $T_\infty$. For the droplets we construct below, we have $T_\infty=0$ and can use conformal symmetry to fix $R_{BH} = 1$ so that we always have $R_{BH}T_\infty=0$, indicative of a droplet phase\footnote{The limit $T_{BH} \to 0$ in our solution does not lead to a funnel as $R_{BH}$ is fixed. Nevertheless, we can see some features of how the Unruh state may settle down to the Hartle-Hawking state at zero temperature.}.

These droplet and funnel configurations are conjectured to correspond, respectively, to the Unruh and Hartle-Hawking vacuum states in the CFT. Typically, these states are characterized by regularity conditions of the stress tensor. The Unruh state is empty at past null infinity and regular on the future horizon whereas the Hartle-Hawking state is regular on both the past and future horizons. There is a third state, the Boulware vacuum, which has an empty stress tensor at both past and future null infinity, and is thus singular at both past and future horizons. The ``detuned" phase of the CFT discussed earlier is thought to correspond to this vacuum. One can also define these vacua by the matter at null infinity. The Hartle-Hawking state has at null infinity a thermal gas in equilibrium with the black hole---hence $T_{BH}=T_\infty$ and this corresponds to the funnel phase. The Unruh state has a flux of outgoing Hawking radiation at the horizon but is empty at null infinity. This suggests $T_\infty=0$ and the black hole acts as a heat source. One would expect in this state that the stress tensor vanishes smoothly as one moves away from the horizon. In our solutions, the stress tensor does in fact vanish as one goes to spatial infinity, but is not monotonic and in $d>4$ even changes sign. As mentioned earlier, the authors of \cite{Figueras:2011va} remain ambivalent over whether the $T_{BH}\neq T_\infty$ droplet is in the Unruh or Boulware state, but suggest $\mathcal{O}(1/N^2)$ effects may point toward to the Boulware vacuum.

The boundary stress tensor is just one avenue for analyzing the state of the boundary field theory. Recently, there has been much excitement over the use of another observable, the entanglement entropy, as a means to characterize quantum field theories. In the context of AdS/CFT, this has been especially exciting because the entanglement entropy of the boundary field theory corresponds to a well-defined geometric quantity in the bulk. On a given time-slice of the field theory background, one may divide the surface into two or more spatial subregions $\{\mathcal{A},\mathcal{B},...\}$. The entanglement entropy of a subregion $\mathcal{A}$ quantifies the entanglement between degrees of freedom in $\mathcal{A}$ and degrees of freedom in its complement $\bar{\mathcal{A}}$. We should emphasize that there is a distinction between the entanglement entropy and von Neumann entropy. In particular, for mixed states such as thermal states of a field theory, the former will vanish when calculated on the whole space while the latter does not. This is because the von Neumann entropy calculates, in addition to the internal correlations of the field theory, correlations between the field theory and the purifying state. For the rest of this paper, we will not distinguish between von Neumann and entanglement entropies. In many cases, especially when the field theory is strongly interacting, the entanglement entropy is difficult to calculate, often requiring the analytic continuation of a path integral on a Riemann surface \cite{Calabrese:2004eu}. Fortunately, for strongly coupled CFTs, we can perform a dual calculation on the gravity side. For static spacetimes, Ryu and Takayanagi \cite{Ryu:2006bv} have conjectured, and Lewkowycz and Maldacena have proven \cite{Lewkowycz:2013nqa}, that the bulk object dual to the entanglement entropy (actually von Neumann entropy) of $\mathcal{A}$ is a co-dimension two minimal surface in the bulk, $\Sigma$, anchored to the conformal boundary at $\partial\mathcal{A}$.\footnote{The extension to stationary spacetimes is given in \cite{Hubeny:2007xt}. First order quantum corrections to this formula were calculated in \cite{Faulkner:2013ana} and extended to all orders in \cite{Engelhardt:2014gca}.} The entanglement entropy in the field theory is then given by the area of this minimal surface in a formula analogous to the Bekenstein-Hawking entropy,
\begin{equation}
S(\mathcal{A}) = \frac{\text{Area}(\Sigma)}{4G_{d+1}}
\end{equation} 
where $G_{d+1}$ is Newton's constant in $d+1$ dimensions.\footnote{It is important to note the RT formula comes with a homology constraint which instructs us to include surfaces that may be disconnected \cite{Headrick:2007km}.} 

The fact that the entanglement entropy is a geometric object in the bulk has inspired many authors to use AdS/CFT to construct bulk spacetimes from knowledge of entanglement in the field theory \cite{Czech:2012bh, Balasubramanian:2013lsa, Faulkner:2013ica, Swingle:2014uza}. Furthermore, it gives an intuitive and visual understanding of entanglement inequalities, renormalization group flow, and confinement-deconfinement phase transitions \cite{Casini:2004bw, Klebanov:2007ws, Hirata:2006jx, Ryu:2006ef, Drukker:1999zq}. While it has been used to understand the properties of thermal field theories on flat backgrounds, studies of entanglement entropy of thermal field theories in black hole backgrounds have been lacking (see \cite{Emparan:2006ni} for early work) and to our knowledge, this is the first work to report the finite, universal terms in this entropy. We hope that these finite terms, as they have in the work on confinement, may bring some new understanding to the problems discussed above and hopefully provided a nice picture of the ``jamming" of the CFT.

Over the last few years, there has been a program of constructing both analytic and numerical funnels and droplets in a journey to understand interacting thermal field theories\cite{Hubeny:2009ru, Hubeny:2009kz, Hubeny:2009rc, Marolf:2013ioa, Figueras:2011va, Figueras:2013jja, Fischetti:2013hja, Fischetti:2012ps, Fischetti:2012vt, Santos:2012he, Santos:2014yja}. Analytic droplets and funnels were constructed in $d=3$ from the AdS C-metric which include an asymptotically flat boundary black hole which will be reproduced below. An analytic funnel dual to the Unruh state was constructed in d=2. Numerical constructions include a d=4 Schwarzschild droplet, $T_{BH} = T_\infty$ funnels, $d=5$ rotating droplets, and $d=2$ ``flowing funnels" in which a detuned CFT phase is seen. One challenge to distinguishing vacuum states is the fact that we have a \emph{conformal} field theory on the boundary. For a d=4 boundary Schwarzschild black hole, we note that we can always rescale the metric such that different Schwarzschild radii, $R_s$, are conformally equal to the $R_s=1$ spacetime. In this case, then, there is no way to vary $T_{BH}$ in a way visible to the CFT. For this reason, we need another parameter on the boundary. The authors of \cite{Fischetti:2013hja} chose to introduce angular momentum to adjust $T_{BH}$. To use the Ryu-Takayanagi method for calculating entanglement entropies, we want our spacetime to be static and so instead, we introduce a ``charge" by imposing Reissner-Nordstr{\"o}m (from here on RN) boundary conditions instead of Schwarzschild. To our knowledge, these droplets have yet to appear in the literature and are therefore new vacuum solutions to the Einstein equations with a negative cosmological constant.

 It should be noted that while RN typically corresponds to a black hole with electric charge, the CFT does not couple to this charge. This is because charged operators on the boundary are dual to charged fields in the bulk. However, we have no matter fields in the bulk and so the CFT has vanishing expectation value for the charge. Thus, the only effect of the charge is to vary the temperature for a fixed value of outer horizon while keeping $T_{\infty}=0$. Interestingly, variation of $T_{BH}$ does affect both the stress tensor expectation value and the entanglement entropy despite $R_{BH}T_\infty =0$.  For numerically constructed solutions with  $d=4, 5, 6$ RN boundary conditions, as well as in the $d=3$ analytic C-metric, the stress tensor and entanglement entropy have universal behavior near spatial infinity that matches the boundary Schwarzschild (or in $d>4$, Tangherlini \cite{Tangherlini:1963bw}) black hole. Near the horizon, however, these observables behave very differently, often including a negative energy density peak that indicates a higher concentration of the jammed plasma. The near horizon behavior is reinforced in the final section where we calculate the entanglement entropy of ball shaped regions on the boundary as a function of both radius and $T_{BH}$ and see interesting behavior at similar locations.

\section{Quantum Stress Tensors in Spherically Symmetric Static Spacetimes}
To understand the numerical results for the boundary stress tensors, we follow the example of \cite{Fischetti:2013hja} and discuss the expectation value of the quantum stress tensor in a static spherically symmetric background. This work extends the analysis of Christensen and Fulling \cite{Christensen:1977jc} to the case of RN in general spacetime dimension d. To keep the field theory arbitrary, we only require the stress tensor be covariantly conserved,
\begin{equation}
\label{eq:conseq}
\nabla_\mu \langle T^{\mu}_{\;\;\nu}\rangle = 0.
\end{equation}
To begin, we work with the following metric
\begin{equation}
ds^2_{RN} = -\Delta_d(R)dt^2+\frac{dr^2}{\Delta_d(R)} + R^2d\Omega_{d-2}^2,\quad\quad \Delta_d(R)=\left(1-\left(\frac{R_+}{R}\right)^{d-3}\right)\left(1-\left(\frac{R_-}{R}\right)^{d-3}\right).
\end{equation}
The most general static spherically symmetric, stress tensor is given by
\begin{equation}
\langle T^{\mu}_{\;\;\nu} \rangle = \left(\begin{array}{ccc} T^t_{\;\;t}&T^{t}_{\;\;R}&0\\T^{R}_{\;\;t}&T^{R}_{\;\;R}&0\\0&0&T^\Omega_{\;\;\Omega}\delta^{i}_{j}\end{array}\right)
\end{equation}
where all components are functions of only $R$ and spherical symmetry tells us that all angular components are equal. Inserting this into (\ref{eq:conseq}), we get the following system of equations:
\begin{equation}
\begin{split}
0&=\partial_RT^{R}_{\;\;t} + \frac{d-2}{R}T^{R}_{\;\;t}\\
0&=\partial_R T^{R}_{\;\;R} +\left(\frac{d-2}{R}-\frac{\Delta_d'(R)}{2\Delta_d(R)}\right)T^{R}_{\;\;R}+\frac{\Delta_d'(R)}{2\Delta_d(R)}T^{t}_{\;\;t}-\frac{d-2}{R}T^{\Omega}_{\;\;\Omega}
\end{split}
\end{equation}
The first equation can be integrated to give
\begin{equation}
T^{R}_{\;\;t} = K(\frac{R_+}{R})^{d-2}.
\end{equation}
where K is an integration constant whose physical importance will be discussed below. Next, we use the trace of the stress tensor to write $T^{R}_{\;\;R}$ in terms of $T^{\Omega}_{\;\;\Omega}$ and $T^{\mu}_{\;\;\mu}$.
\begin{equation}
T^{R}_{\;\;R} = \frac{(R_+/R)^{d-2}}{\Delta_{d}(R)}\left[Q-K+\frac{1}{2}\int_{R_+}^{R}(\tilde{R}/R_+)^{d-3}\left(\tilde{R}\Delta_d'T^{\mu}_{\;\;\mu}+(d-2)(2-\tilde{R}\Delta_d')T^{\Omega}_{\;\;\Omega}\right)\frac{d\tilde{R}}{R_+}\right]
\end{equation}
where Q is another integration constant to be discussed below. It will be helpful to split the stress tensor into four terms
\begin{equation}
T^{\mu}_{\;\;\nu} = (T_1)^\mu_{\;\;\nu} +(T_2)^\mu_{\;\;\nu} +(T_3)^\mu_{\;\;\nu} +(T_4)^\mu_{\;\;\nu}
\end{equation}
The first term contains only information about the trace,
\begin{equation}
(T_1)^\mu_{\;\;\nu} = \text{diag}\left\{ -\frac{(R_+/R)^{d-2}}{\Delta_d(R)}H(R) + \frac{1}{2}T^{\mu}_{\;\;\mu}(R),  \frac{(R_+/R)^{d-2}}{\Delta_d(R)}H(R), \frac{1}{2(d-2)}T^{\mu}_{\;\;\mu}(R)\delta^{i}_{j}\right\}
\end{equation}
where 
\begin{equation}
H(R) \equiv \frac{1}{2}\int_{R_+}^{R}(\frac{\tilde{R}}{R_+})^{d-3}(\tilde{R}\Delta_d' - \Delta_d)T^{\mu}_{\;\;\mu}(\tilde{R})\frac{d\tilde{R}}{R_+}.
\end{equation}
Note that we only construct exteriors of black holes and so the integration is only for $R\geq R_+$. In odd boundary dimensions or Ricci flat spacetimes, there is no conformal anomaly and so the trace of the stress tensor vanishes. However, we construct solutions in both even and odd boundary dimensions which are not Ricci flat and $(T^1)^{\mu}_{\;\;\nu}$ can contribute. 

The next term in the stress tensor tells us that the flux of Hawking radiation at null infinity is proportional to $K$. In terms of the tortoise coordinate $dR_*=dR/\Delta_d(R)$, 
\begin{equation}
(T_2)^{\mu}_{\;\;\nu} = K\frac{(R_+/R)^{d-2}}{\Delta_d(R)}\left(\begin{array}{ccc}1&1&0\\-1&-1&0\\0&0&0\end{array}\right)
\end{equation}
Since we construct solutions with no heat transfer at infinity, we expect that $K=0$ for our solutions. In particular, our stress tensor should fall off faster than $R^{2-d}$. Below, we will  see that our stress tensors fall off as $R^{-(d+1)}$ satisfying this criteria. The third term is proportional to $Q$ and tells us about regularity at the future horizon,
\begin{equation}
(T_3)^{\mu}_{\;\;\nu} = Q \frac{(R_+/R)^{d-2}}{\Delta_d(R)}\text{diag}\{-1,1,0\}
\end{equation}
In particular, the diverging denominator tells us regularity on this horizon requires that $Q=0$. Finally, we have a term that determines the pressures. Defining the functions
\begin{equation}
\begin{split}
\Theta(R) &\equiv T^{\Omega}_{\;\;\Omega}(R) - \frac{1}{2(d-2)}T^{\mu}_{\;\;\mu}(r)\\
G(R) &\equiv \frac{d-2}{2}\int_{R_+}^{R}\left((\frac{\tilde{R}}{R_+})^{d-3}(2-\tilde{R}\Delta_d')\right)\Theta(\tilde{R})\frac{d\tilde{R}}{R_+},
\end{split}
\end{equation}
we may write 
\begin{equation}
(T_4)^{\mu}_{\;\;\nu}=\text{diag}\left\{-\frac{(R_+/R)^{d-2}}{\Delta_d(R)}G(R)-(d-2)\Theta(R),\frac{(R_+/R)^{d-2}}{\Delta_d(R)}G(R),\Theta(R)\delta^{i}_{j}\right\}
\end{equation}
For Ricci flat spacetimes regular on both horizons, this is the only part of the stress tensor that is non-vanishing. Because our approach was completely general, any spherically symmetric static quantum stress tensor will have this form, including the strongly interacting one that we consider below. By matching onto this solution, we can draw conclusions about the nature of our jammed CFT. In particular, Christensen and Fulling suggest that states with $Q=K=0$ which have regular horizons and no flux at null infinity are dual to the Unruh state. 

As pointed out by Fischetti and Santos, the notion of single particle states in field theories become ambiguous when put on curved backgrounds. However, currents such as the stress tensor remain well defined even in the presence of background curvature. External fields like the curvature may couple to these currents and can lead to interesting new behavior like the conformal anomaly. One peculiar feature of the stress tensor in black hole backgrounds that our results exhibit is a negative energy density. From free field theory in Minkowski spacetime, this may seem paradoxical, but as Fischetti and Santos point out, even there, a negative local energy density appears in the Casimir effect. Furthermore, they emphasize that this negative energy density seems to be typical of free field theories near black hole horizons in both the Unruh and Hartle-Hawking states \cite{Epstein:1965zza, Davies:1977yv}. This, they say, is consistent with the picture of Hawking radiation as pair-production with negative energy particles falling into the black hole and positive energy particles escaping. While the particle-antiparticle picture may not apply to our strongly interacting field theory, we still expect that this negative energy density should be ubiquitous as our results below confirm. Contrary to their results however, this energy density becomes positive away from the black hole horizon for $R_-/R_+$ sufficiently large. Interestingly, in $d=6$, we see that the region of negative energy density becomes disconnected from the horizon as the black hole nears extremality.


\section{Numerical Construction of RN Boundary Black Holes}
In order to construct the background spacetimes for our field theory, we solve the DeTurck equations with a negative cosmological constant in d+1 dimensions,
\begin{equation}
R_{AB} - \frac{2\Lambda}{d-1}g_{AB}-\nabla_{(A}\xi_{B)}=0,\quad\quad \xi^A = g^{BC}(\Gamma^A_{BC}-\bar{\Gamma}^{A}_{BC})
\label{eq:DeTurck}
\end{equation}
with $2\Lambda = -(d-1)(d)/L_{AdS_{d+1}}^2$. In this expression, we have introduced Latin letters to denote bulk spacetime indices. The DeTurck vector, $\xi^A$, is defined in terms of a Levi-Civita connection, $\bar{\Gamma}^{A}_{BC}$, derived from a reference metric of our choice $\bar{g}$. Equation (\ref{eq:DeTurck}) is a deformation of the Einstein field equations which, when evaluated on a solution with $\xi^A=0$, is analogous to a choice of gauge. As was shown by the authors of \cite{Figueras:2011va}, this deformation gives an elliptic differential equation which is better suited to numerical evaluation. Furthermore, these authors showed that given a stationary spacetime with Killing horizons, the maximum of $\xi^2=\xi^A\xi_A$ must occur at the boundaries (or ``fictitious boundaries" like symmetry axes and black hole horizons). With a suitable choice of reference metric, $\bar{g}$, that has $\xi^A=0$ on the boundaries, solutions to the DeTurck equations should also be solutions to the Einstein equations. To confirm this, we monitored the magnitude of $\xi^2$ and we check that once obtained, our solutions satisfy the Einstein equations to the same precision. Our construction of boundary AdS/RN black holes will closely follow \cite{Figueras:2011va} who constructed a five dimensional droplet solution corresponding to a four dimensional boundary Schwarzschild black hole with an extremal bulk horizon at $T_{\infty}=0$.

We would like to construct static, asymptotically Anti-de Sitter, spherically symmetric solutions corresponding to an asymptotic field theory plasma at $T_\infty = 0$. From the AdS/CFT correspondence, this tells us that we need a bulk black hole which has an asymptotic planar black hole at $T_\infty = 0$. This is an extremal horizon and we know that this must correspond to the IR horizon of Poincar{\'e}-AdS. This horizon is at $z\to \infty$ and so to construct it numerically, we must choose a new AdS radial coordinate. We start with pure Poincar{\'e} AdS in d+1 dimensions,\footnote{Note that we have chosen to use Euclidean time, although because our solution is static, we could just as easily construct Lorentzian solutions. Because we will evaluate the stress tensor with one index up and one index down, i.e. $\langle T^{\mu}_{\;\;\nu}\rangle$, this choice of time coordinate will give the same results as for the Lorentzian analysis above.}

\begin{equation}
ds^2 = \frac{l^2}{z^2}(dz^2 + d\tau^2 + dR^2 + R^2 d\Omega_{d-2}^2).
\end{equation}
 Next, we make the coordinate change,
\begin{equation}
\label{eq:pureAdScoordinates}
R = \frac{x\sqrt{2-x^2}}{1-r^2},\quad z=\frac{1-x^2}{1-r^2},\quad 0\leq x \leq 1, 0\leq r <1
\end{equation}
so that the metric becomes
\begin{equation}
ds^2 = \frac{l^2}{(1-x^2)^2}\left(f(r)^2 d\tau^2 + \frac{4r^2}{f(r)^2}dr^2 + \frac{4}{g(x)}dx^2 + x^2g(x)d\Omega_{d-2}^2\right)
\end{equation}
where 
\begin{equation}
f(r) = 1-r^2, \quad g(x) = 2-x^2.
\label{eq:deffandg}
\end{equation}

In these coordinates, the conformal boundary is located at $x=1$, while the Poincar{\'e} horizon is located at $r=1$. The axis of rotational symmetry is at $x=0$.  

We would like to deform this solution in such a way that the conformal boundary has the form of a d-dimensional Reissner-Nordstr{\"o}m black hole. Since we want this to be a droplet, this horizon extends into the bulk and smoothly ends at the symmetry axis. If we define $r=0$ as the horizon location, then the following metric ansatz will have such a horizon,
\begin{equation}
\label{eq:ansatz}
ds^2 = \frac{(1-r^2)^2}{(1-x^2)^2}\left(r^2Td\tau^2 + \frac{4A}{f(r)^4}dr^2 + \frac{4B}{f(r)^2g(x)}dx^2 + \frac{2rxF}{f(r)^3}drdx + \frac{x^2g(x)S}{f(r)^2}d\Omega_{d-2}^2\right)
\end{equation}
where $X\equiv\{T,S,A,B,F\}$ are all functions of $x$ and $r$. Note that smoothness of the metric functions X tells us that pure AdS$_d$ is not within our ansatz as this would require $T=1/r^2$. 
 
We require that our spacetime, as $x\to 1$, is asymptotically locally AdS with a metric conformal to d-dimensional Reissner-Nordstr{\"o}m. In the limit $x \to 1$, (\ref{eq:pureAdScoordinates}) becomes 
$R=1/1-r^2$. We want this to have dimensions of length, and so we define for our boundary metric,
\begin{equation}
\label{eq:boundaryRdefinition}
R = \frac{R_+}{1-r^2}.
\end{equation}
As before, the d-dimensional Reissner-Nordstr{\"o}m metric is,
\begin{equation}
\begin{split}
ds^2 &=\Delta_d(R) d\tau^2+ \frac{dR^2}{\Delta_d(R)} + R^2 d\Omega_{d-2}^2\\
\Delta_d &= 1-\frac{2M_d}{R^{d-3}} + \frac{Q_d^2}{R^{2(d-3)}} = \left(1-(\frac{R_+}{R})^{d-3}\right)\left(1-(\frac{R_-}{R})^{d-3}\right)\\
\label{eq:Deltadofr}
\end{split}
\end{equation}
where $R_{\pm}^{d-3} = M_d\pm \sqrt{M_d^2-Q_d^2}$ and $M_d,Q_d$ are related to the energy, $\mu$ and charge, $q$ of the d-dimensional black hole in the following way \cite{Horowitz:2012nnc},\footnote{The charge comes from considering an electric field $E = q/\Omega_{d-2}r^{d-2}$.}
\begin{equation}
M_d = \frac{16\pi G_d}{(d-2)\Omega_{d-2}}\mu\quad\quad Q_d^2 = \frac{8\pi G_d}{(d-3)(d-2)}\frac{q^2}{\Omega_{d-2}^2}.
\end{equation}
These black holes have temperatures (in natural units)
\begin{equation}
T_d= \frac{\kappa_d}{2\pi}\quad\text{where}\;\;\kappa_d = \frac{(d-3)\left(1-\left(\frac{R_-}{R_+}\right)^{d-3}\right)}{2R_+}
\end{equation}
After the change of variables the boundary metric becomes
\begin{equation}
\label{eq:boundarymetric}
ds^2 = g_{\mu\nu}dx^\mu dx^\nu= r^2\delta_d(r)dt^2 + \frac{4R_+^2dr^2}{(1-r^2)^4\delta_d(r)} + \frac{R_+^2}{(1-r^2)^2}d\Omega_{d-2}^2
\end{equation}
where
\begin{equation}
\delta_d(r) = \frac{1}{r^2}\left(1-(1-r^2)^{d-3}\right)\left(1-(1-r^2)^{d-3}\left(\frac{R_-}{R_+}\right)^{d-3}\right)
\end{equation}
Near the boundary, we want
\begin{equation}
ds^2 \to \frac{(1-r^2)^2}{(1-x^2)^2}\left(\frac{1}{f(r)^2}dx^2 + g_{\mu\nu}dx^\mu dx^\nu\right)
\end{equation}
Now, we set $l=1$ and use conformal symmetry to fix $R_+ = 1$. In particular, note that in the limit $R_-\to 0$, we can take $\tau\to R_+\tau$ and the parameter $R_+$ completely scales out of the metric (\ref{eq:boundarymetric}). This means that, to the conformally invariant theory, all boundary Schwarzschild black holes are equivalent---hence our need for another parameter, $R_-$. When we set $R_+=1$, we choose a particular branch of RN solutions such that
\begin{equation}
\kappa_d = (d-3)(1-M_d) = \frac{(d-3)(1-Q_d^2)}{2}
\end{equation}
Following the above discussion, we impose the following boundary conditions on $X$. As $x\to 1$,
\begin{equation} 
\label{eq:conformalBCs}
 T\to \delta_d(R),\;S\to 1,\;A\to \frac{1}{\delta_{d}(r)},\;B\to 1,\; F\to 0.
\end{equation}
As $r\to 1$,
\begin{equation}
\begin{split}
 T\to 1+T_1(1-r),\;S\to 1+S_1(1&-r),\;A\to 1 +A_1(1-r),\; B\to 1+B_1(1-r),\; F \to (1-r)F_1,\\
&(T_1 - A_1)_{r=1} = \text{constant}.
\end{split}
\end{equation}
The last boundary condition is required to ensure this boundary is an extremal horizon. 

The rotation axis and droplet horizon serve as fictitious boundaries. Figueras et al. show that the DeTurck problem is still well defined on these fictitious boundaries. For these boundaries, we require that our solutions be smooth---as we approach the horizon $r\to 0$, $X$ must be functions of $r^2$, and as $x\to 0$, $X$ must depend only on $x^2$ so that 
\begin{equation}
\partial_rX|_{r=0}=0, \quad \partial_xX|_{x=0}=0.
\end{equation}
Furthermore regularity of the Euclidean solution at the horizon requires
\begin{equation}
\frac{T}{A}|_{r=0} = 4\kappa_d^2 
\end{equation}
and regularity at the rotation axis requires
\begin{equation}
\frac{S}{A}|_{x=0} = 1.
\end{equation}
Finally, our choice of reference metric, $\bar{g}_{AB}$ is (\ref{eq:ansatz}) with
\begin{equation}
T = \frac{1}{\delta_d(r)},\;\; A = \delta_d(r),\;\; S=B=1, \;\; F=0.
\end{equation}

\begin{figure}[t!]
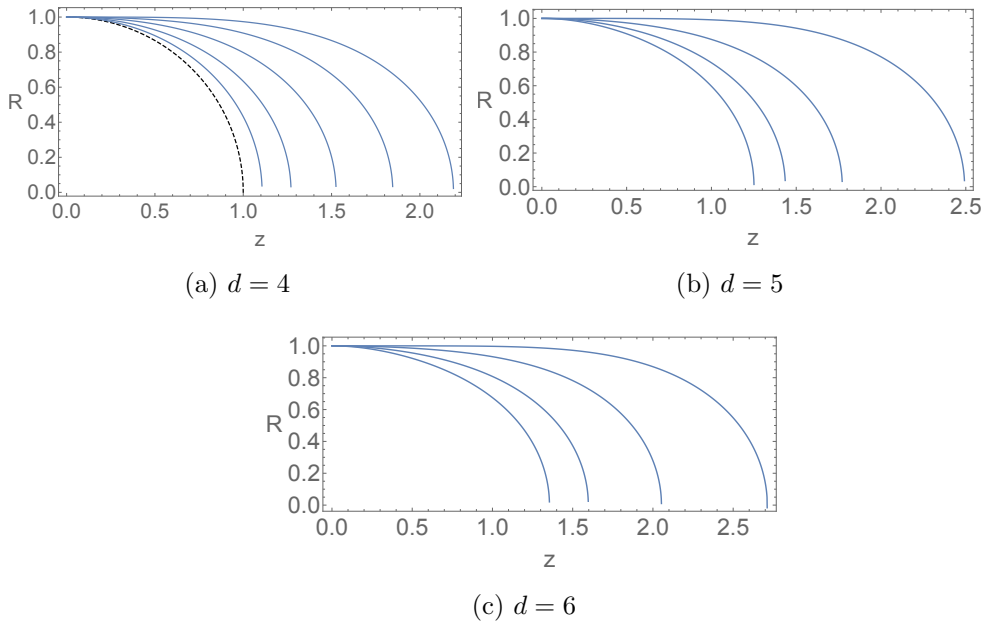

\centering
\begin{subfigure}[b]{.4\textwidth}
\includegraphics[width=\textwidth]{5dembedding.png}
\caption{$d=4$}
\end{subfigure}
\begin{subfigure}[b]{.44\textwidth}
\includegraphics[width=\textwidth]{6dembedding.png}
\caption{$d=5$}
\end{subfigure}
\newline
\begin{subfigure}[b]{.45\textwidth}
\includegraphics[width=\textwidth]{7dembedding.png}
\caption{$d=6$}
\end{subfigure}
\caption{In (a)-(c), we embed the horizons into the metric $ds^2 = \frac{l^2}{z^2}(dz^2+dR^2+R^2d\Omega_{d-2}^2)$. In (a), we choose (from right to left) $R_-/R_+ = 0,\;.2,\;.5,\;.8,\;.96,\;1.0$. The extremal black hole is the dotted black line. In (b) and (c), we choose (R to L) $R_-/R_+ =0,\;.5,\;.7,\;.8$. Notably, the extremal horizon in $d=4$ is a minimal surface in the pure AdS where it is embedded.}
\label{fig:embeddings}
\end{figure}

\begin{figure}[t]
\centering
\begin{subfigure}[b]{.48\textwidth}
\includegraphics[width=\textwidth]{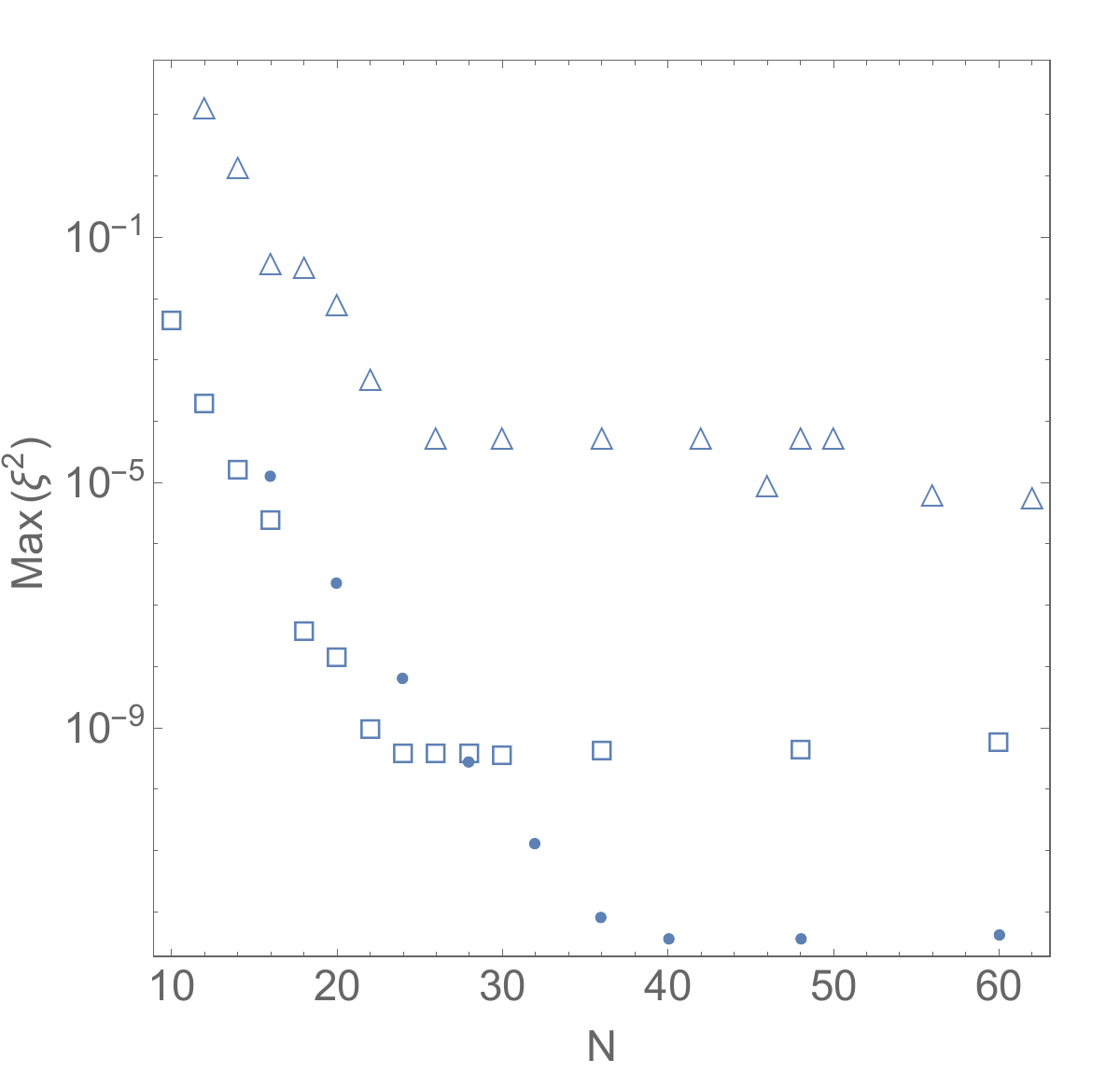}
\caption{$R_-/R_+ = .2$}
\end{subfigure}
\begin{subfigure}[b]{.48\textwidth}
\includegraphics[width=\textwidth]{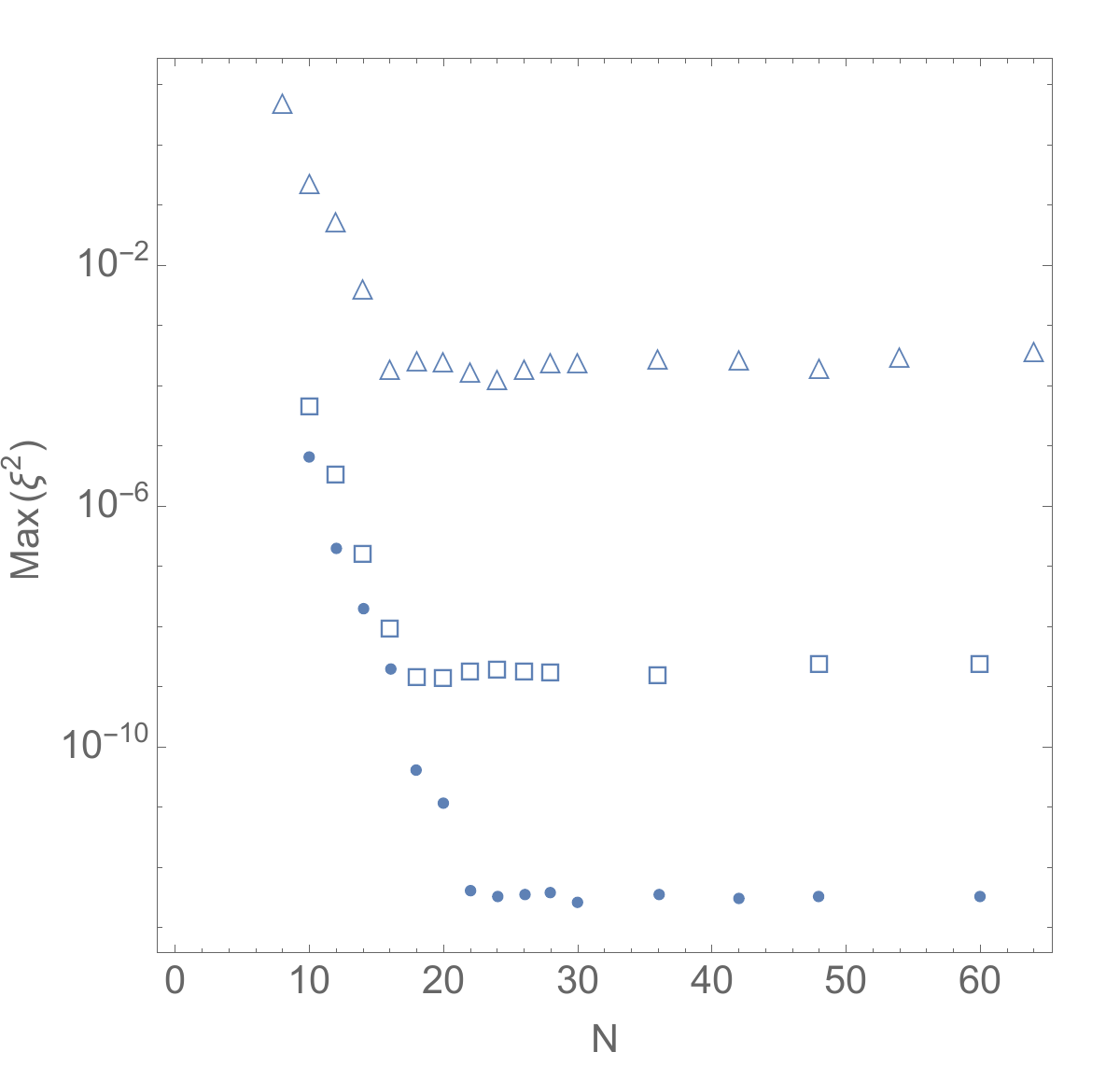}
\caption{$R_-/R_+=.5$}
\end{subfigure}
\caption{\label{fig:convergence} The maximum value of $\xi^2$ for two choices of $R_-/R_+$ in $d=4$ (circles), $d=5$ (squares), and $d=6$ (triangles). Our numerical method leads to exponential convergence in the number of grid points, $N$, until saturation. Even boundary dimensions ($d=$ 4 and 6) show more numerical error because of the presence of logarithmic terms in the asymptotic expansion (see equation (\ref{eq:asymptoticexpansions}).)}
\end{figure}

To find solutions, we use a Newton-Raphson relaxation algorithm using pseudospectral collocation on a Chebyshev grid. In $d=5$, we found $\xi^2 \sim 10^{-13}$ for all solutions below. In $d= 4$ and $6$, the numerics are slightly more unstable, and more grid points were necessary. For the largest grids we used, $81\times81$ in $d=6$ we found $\xi^2\sim 10^{-10}$ in $d=4$ and $\xi^2 \sim 10^{-6}$ in $d=6$. Plots of convergence for two characteristic choices of $R_-$ are shown in figure \ref{fig:convergence}. In fig. \ref{fig:embeddings}, we plot the droplet horizons of our solutions by embedding them in the metric $ds^2 = \frac{l^2}{z^2}(dz^2 + dR^2 +R^2d\Omega_{d-2}^2)$. In four dimensions, for $R_-=R_+$, the geometry on the horizon exactly matches the surface $R^2+z^2 = R_+^2$. Importantly, this is a minimal surface in pure AdS, as we will discuss below. In higher dimensions, the horizon approaches this surface, but sufficiently close to extremality, the horizon can no longer be isometrically embedded. The largest $R_-$ we plot is approximately this critical value. 

\section{Boundary Stress Tensor}

As discussed above, our spacetime is asymptotically locally Anti-de Sitter. This means the metric can be expanded in a neighborhood of the boundary in terms of the Fefferman-Graham coordinate, $z$ \cite{Fischetti:2012rd}. The boundary stress tensor can be determined from the coefficients of $z^{i}$ for $i\leq d$. The expansion and expressions for the boundary stress tensor in terms of these coefficients is discussed in the appendix. For the boundary stress tensor of our numerical solutions, we need to find an expression for the coordinate $z$ and boundary radial coordinate $R$ in terms of $x$ and $r$ as well as boundary expansions for the functions $X$. To do so, we write
\begin{equation}
\begin{split}
\label{eq:asymptoticexpansions}
z &= (1-x^2)\left(\frac{1}{1-r^2} + \sum_{n=1}^{\infty} z_{n}(r)(1-x^2)^{n}\right)\\
R&= \frac{R_+}{1-r^2} + \sum_{n=1}^{\infty} R_{n}(r)(1-x^2)^{n}\\
X&=X_0(r) +  \sum_{n=1}^{\infty} X_{n}(r)(1-x^2)^{n} + \log(1-x^2)\sum_{n=1}^{\infty} \tilde{X}_n(r)(1-x^2)^{n}
\end{split}
\end{equation}
where $X_0(r)$ are our Dirichlet boundary conditions (\ref{eq:conformalBCs}). By inserting the expansion for $X$ into the DeTurck equations and matching with the Fefferman-Graham expansion, we can find the functions $z_n, R_n$ and $X_n$ in terms of $x,r$ for all $n<d$ including potential log$(1-x^2)^d$ terms in even $d$. In the appendix, we present the functions $X_n$ relevant to calculating the boundary stress tensor. Importantly, these terms are also sufficient to determine the UV divergences in the entanglement entropy.

\begin{figure}[t]
\centering
\begin{subfigure}[b]{.32\textwidth}
\includegraphics[width=\textwidth]{5dinterpolationerrors.png}
\caption{$d=4$}
\end{subfigure}
\begin{subfigure}[b]{.32\textwidth}
\includegraphics[width=\textwidth]{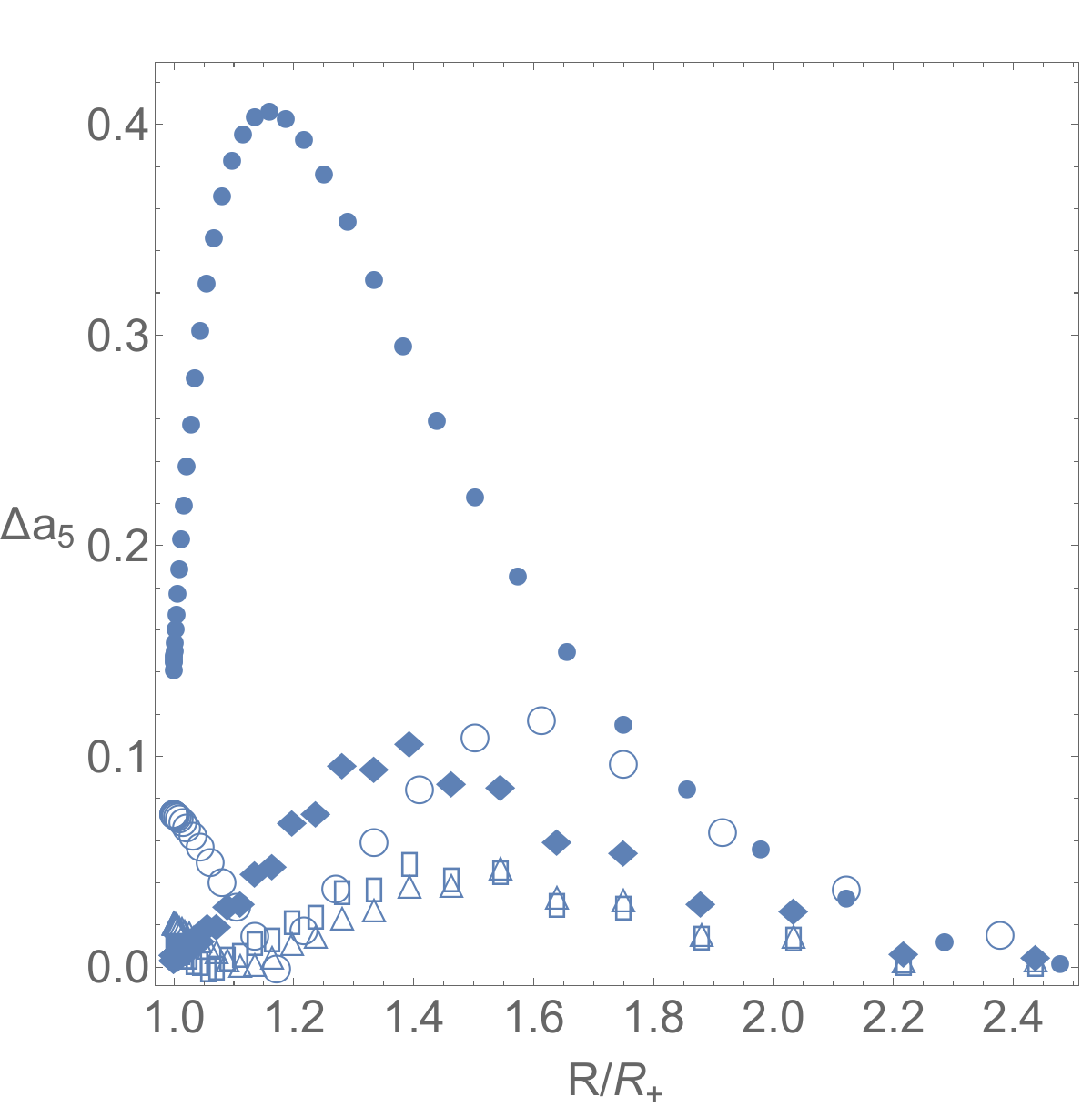}
\caption{$d=5$}
\end{subfigure}
\begin{subfigure}[b]{.32\textwidth}
\includegraphics[width=\textwidth]{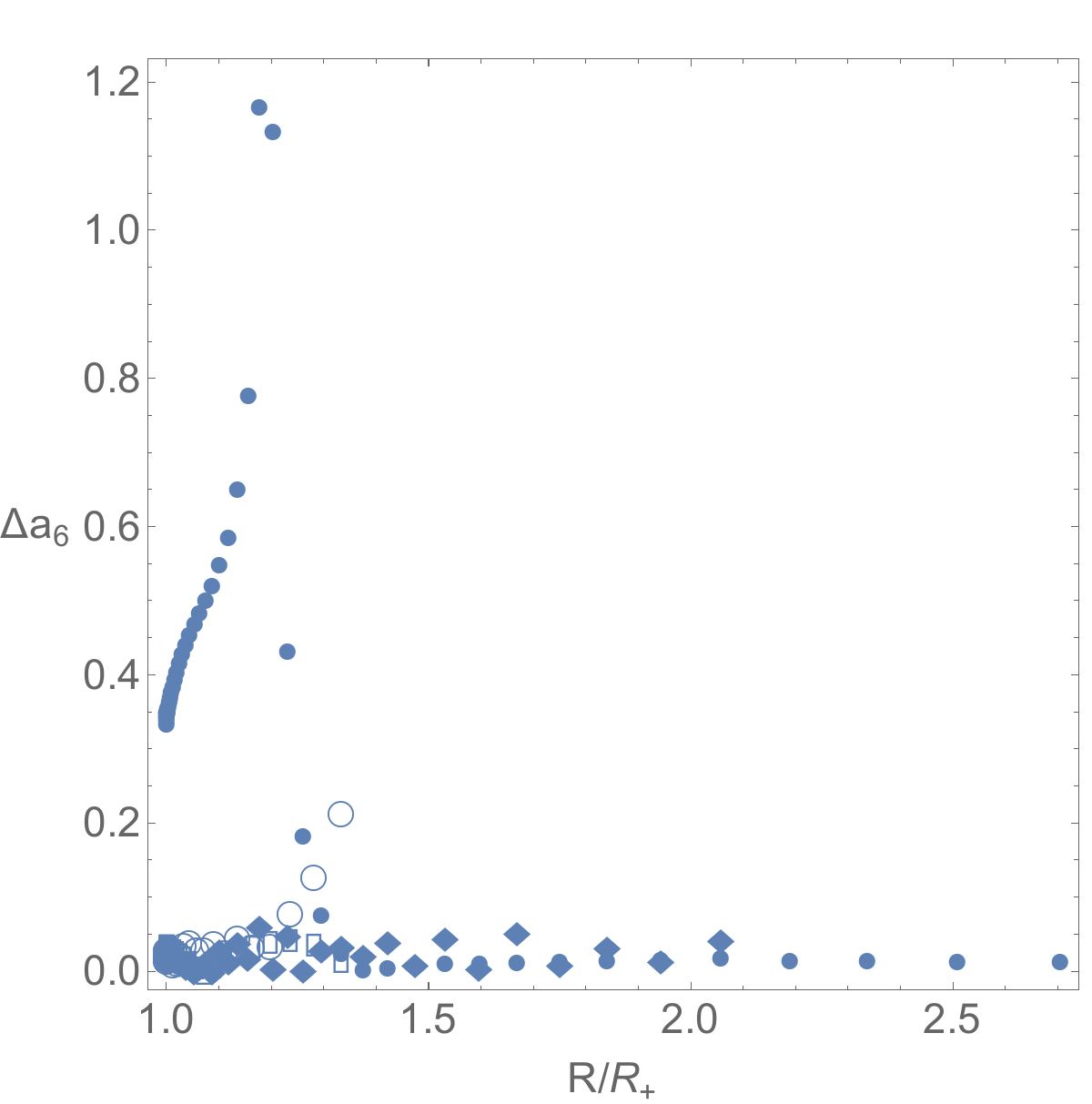}
\caption{$d=6$}
\end{subfigure}
\caption{In (a)-(c), we plot the error in extracting $A_d(r)$. Regions with large errors (especially in (c)) correspond to places where our calculated $A_d(r)$ vanishes while our theoretical $\mathcal{A}_d(r)$, while small, does not vanish exactly. Away from these points, the errors are a few percent or less. Furthermore, in $d=6$ as seen in (c), because of large coefficients, errors accumulate quickly. In each plot, the different values of $R_-/R_+$ match the values used for the energy densities in figs. \ref{fig:5dstress}, \ref{fig:6dstress}, and \ref{fig:7dstress}. In order of increasing $R_-/R_+$, the symbols are $\bullet ,\blacklozenge ,\triangle ,\square,\bigcirc$. (For $d=6$, we don't use $\square$.)}
\label{fig:interpolationerrors}
\end{figure}
The $X_d(r)$ terms are relevant to the boundary stress tensor and must be found numerically. Unfortunately, high order derivatives are numerically unstable, and so to find the coefficients $X_d(r)$, we subtract the known expressions for $X_n$ with $n<d$ above from our numerical solution and fit this to the $X_d(r)$ term in the expansion near the boundary,
\begin{equation}
X^{\text{num}}(x,r)-\sum_{n<d}X_{n}(1-x^2)^n = (1-x^2)^d X_d(r).
\end{equation}
To monitor the numerical accuracy of this method, we note that, because the trace of the stress tensor is known, one of our coefficients can be calculated from knowledge of the other coefficients. We chose to specifically monitor the function $A_d(r)$. The analytic expression for this function, which we call $\mathcal{A}_d(r)$, in terms of the other $X_d(r)$ can be found in the appendix. In fig. \ref{fig:interpolationerrors}, we plot 
\begin{equation}
\Delta a_d(r) \equiv \left|\frac{\mathcal{A}_d(r)-A_d(r)}{A_d(r)}\right|
\label{eq:interpolationdiff}
\end{equation}
for the values of $R_-/R_+$ that we display in the stress tensors below. In $d=4$, the errors stay below a few percent for all $R$ and most are less than a percent. In higher dimensions the errors increase, especially close to the horizon. These errors are due, in most cases, to the stress tensor changing sign. If $\mathcal{A}_d$ and $A_d$ cross the axis at different values of $R$, the denominator of (\ref{eq:interpolationdiff}) blows up. Away from these locations, the errors again become on the order of a few percent or less. For $R_-/R_+=0$ in $d=6$, there is an error close to the horizon where the stress tensor does not vanish. Instead, this can be traced to the large coefficients in $\mathcal{A}_6(r)$ (\ref{eq:Ad}) which cause errors to accumulate quickly. For this case, we checked that the stress tensor does not change appreciably as we varied the grid size.

\begin{figure}[t]
\centering
\begin{subfigure}[b]{.45\textwidth}
\includegraphics[width=\textwidth]{5dRmpoint0StressTensor.png}
\caption{$R_-/R_+ = 0$}
\end{subfigure}
\begin{subfigure}[b]{.45\textwidth}
\includegraphics[width=\textwidth]{5dRmpoint5StressTensor.png}
\caption{$R_-/R_+=.5$}
\end{subfigure}
\newline
\begin{subfigure}[b]{.45\textwidth}
\includegraphics[width=\textwidth]{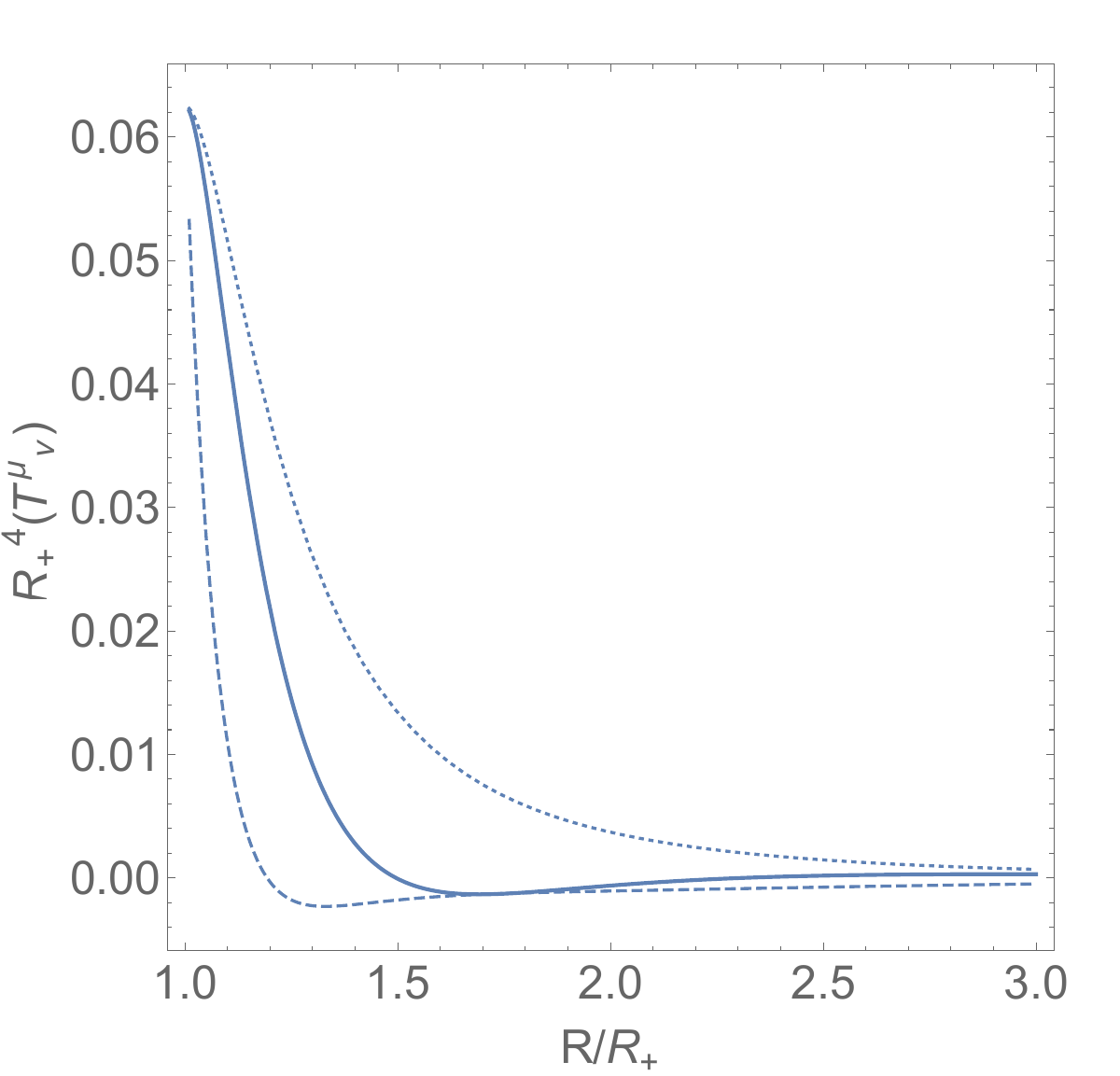}
\caption{$R_-/R_+ = 1.0$}
\end{subfigure}
\begin{subfigure}[b]{.45\textwidth}
\includegraphics[width=\textwidth]{5denergydensities.png}
\caption{}
\label{fig:5denergydensities}
\end{subfigure}
\caption{(a)-(c) are plots of the four dimensional $\frac{4\pi G_5}{l^3}\langle T^{\mu}_{\;\;\nu}\rangle$ as a function of $R$ for different values of $R_-/R_+$. In the plots, the thick line is the energy density $T^{t}_{\;\;t}$, the dotted line is $T^{r}_{\;\;r}$, and the dashed line is $T^{\Omega}_{\;\;\Omega}$. In (d), we plot the energy densities for $R_-/R_+ =$ 0, .2, .5, .9, and 1.0 (black dotted line).}
\label{fig:5dstress}
\end{figure}
$\newline$
Finally, we write the expressions for the boundary stress tensors. In d=4,
\begin{equation}
\langle T^{\mu}_{\;\; \nu} \rangle= \frac{l^3}{4\pi G_5}\text{diag}\biggl\{T^t_{\;\;t},T^{R}_{\;\;R},T^{\Omega}_{\;\;\Omega},T^{\Omega}_{\;\;\Omega}\biggr\}
\end{equation}
where
\begin{equation}
\begin{split}
T^t_{\;\;t}&=\frac{1}{R^4}\biggl(\frac{T_4(R)}{(1-\frac{R_-}{R})}+\frac{3R_+}{4R}(1-\frac{R_+}{R})\\&\quad\quad\quad\quad\quad\quad\quad+  R_- \frac{ \left(((78R_+-12 R) R-73R_+^2) R_-+ R (R (12 R-82R_+)+78R_+^2)R_+\right)}{16 R^4}\biggr)
\end{split}
\end{equation}
\newline
\begin{equation}
\begin{split}
T^{R}_{\;\;R}&=\frac{1}{R^4}\left(\frac{3R_+^2}{4R^2} - (\frac{T_4(R)}{1-\frac{R_-}{R}}+2S_4(R))+R_-\frac{ \left((2 R (6 R-17R_+)+35R_+) R_-+2 R (5 R-17R_+) R_+\right)}{16 R^4}\right)
\end{split}
\end{equation}
and
\begin{equation}
\begin{split}
T^{\Omega}_{\;\;\Omega}&=\frac{1}{R^4}\left(-\frac{3R_+}{8R}+S_4(R)+R_-\frac{ \left((21R_+-22 R) R_+R_--2 R (3 (R-6R_+) R+11R_+^2)\right)}{16 R^4}\right)
\end{split}
\end{equation}
Notably, as $R_-\to 0$, this agrees with Figueras et al. Furthermore the trace is 
\begin{equation}
\langle T^{\mu}_{\mu} \rangle = \frac{l^3}{4\pi G_5}\frac{R_+^2R_-^2}{4R^8}
\end{equation}
agreeing with the conformal anomaly in 4 dimensions.

\begin{figure}[t]
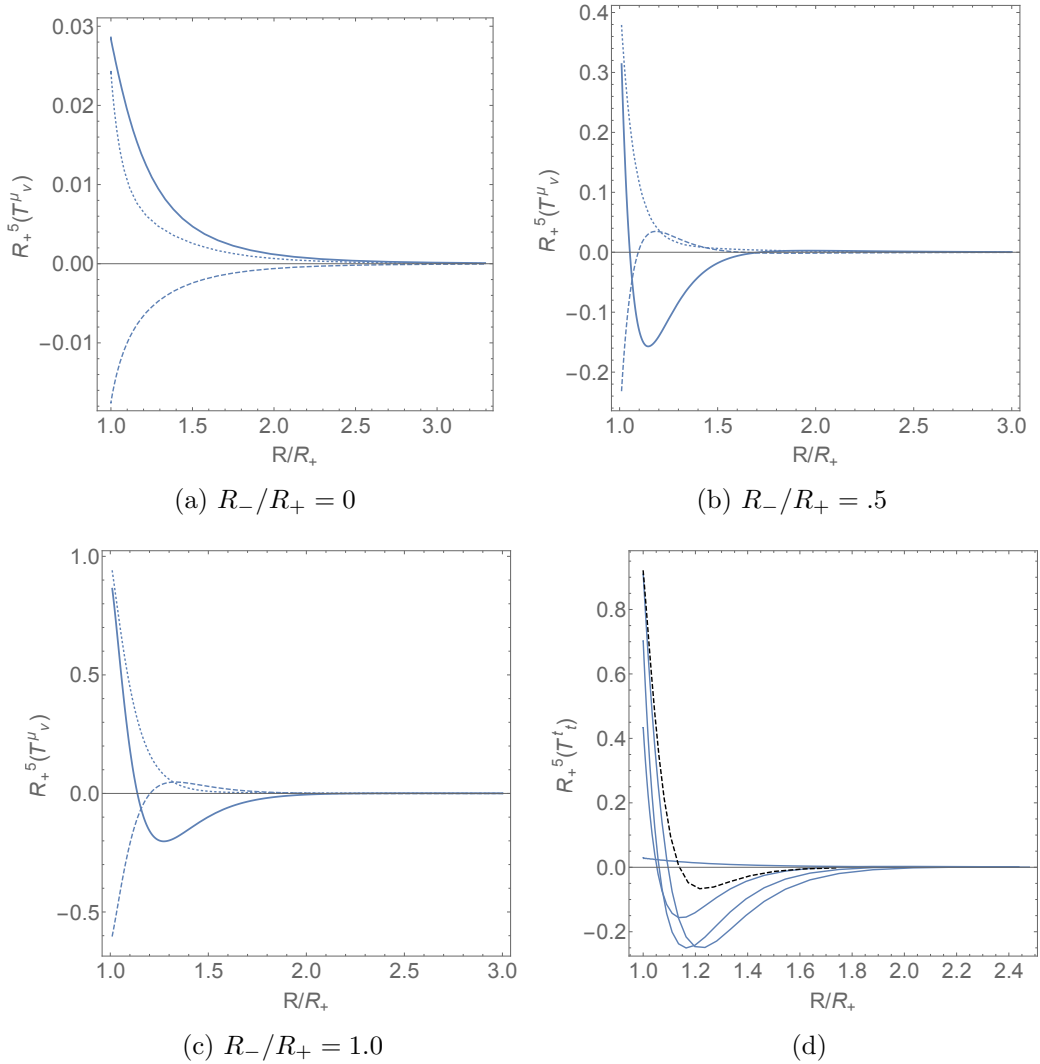

\centering
\begin{subfigure}[b]{.45\textwidth}
\includegraphics[width=\textwidth]{6dRmpoint0StressTensor.png}
\caption{$R_-/R_+ = 0$}
\end{subfigure}
\begin{subfigure}[b]{.45\textwidth}
\includegraphics[width=\textwidth]{6dRmpoint5StressTensor.png}
\caption{$R_-/R_+=.5$}
\end{subfigure}
\newline
\begin{subfigure}[b]{.45\textwidth}
\includegraphics[width=\textwidth]{6dRm1point0StressTensor.png}
\caption{$R_-/R_+ = 1.0$}
\end{subfigure}
\begin{subfigure}[b]{.45\textwidth}
\includegraphics[width=\textwidth]{6denergydensities.png}
\caption{}
\label{fig:6denergydensities}
\end{subfigure}
\caption{(a)-(c) are plots of the five dimensional $\frac{16\pi G_5}{5l^4}\langle T^{\mu}_{\;\;\nu}\rangle$ as a function of $R$ for different values of $R_-/R_+$. In the plots, the thick line is the energy density $T^{t}_{\;\;t}$, the dotted line is $T^{r}_{\;\;r}$, and the dashed line is $T^{\Omega}_{\;\;\Omega}$. In (d), we plot the energy densities for $R_-/R_+ = $0, .5, .7, .9, and 1.0 (black, dotted line).}
\label{fig:6dstress}
\end{figure}

In fig. \ref{fig:5dstress}, we plot this stress tensor for different values of $R_-$. This stress tensor agrees with results of \cite{Figueras:2011va} in the limit $R_- \to 0$. In this limit, it is clear that the trace of the stress tensor vanishes. Furthermore, in the language of section 2, $Q=K=H=0$ indicative of the Unruh vacuum. Interestingly, we see new behavior in the CFT  as $R_-/R_+$ increases. As mentioned earlier, for all $R_-$, the stress tensor displays negative energy densities near the horizon\footnote{Recall that in both Euclidean and Lorentzian signature $\langle T^t_{\;\;t}\rangle<0$ would indicate positive energy.}. This is typical of the non-classical state we expect from a strongly interacting field theory. On the other hand, we see that as we approach extremality, there is a turning point in the energy density. Furthermore, as this ratio becomes sufficiently large, there is a finite size region near the horizon with positive energy density. In droplets with $T_\infty>0$, the authors of \cite{Santos:2014yja} saw positive energy densities in this same limit. In this limit, the pressure also becomes positive near the horizon but becomes negative far away, matching the $R_-=0$ behavior. Finally, as seen in fig. \ref{fig:5denergydensities}, the magnitude of the energy density near the horizon actually decreases as the boundary black hole approached extremality. This is different than what was observed in \cite{Fischetti:2013hja} where the magnitude of the energy density increased monotonically as the black hole approached extremality. We believe this behavior may indicate that the plasma is becoming localized away from the horizon as the Unruh and Hartle-Hawking states degenerate at zero temperature. We propose that the peak in the energy density corresponds roughly to the location of the jammed CFT. This is reinforced by the entanglement entropy calculations.
 
\begin{figure}[t]
\centering
\begin{subfigure}[b]{.45\textwidth}
\includegraphics[width=\textwidth]{7dRmpoint0StressTensor.png}
\caption{$R_-/R_+=0$}
\end{subfigure}
\begin{subfigure}[b]{.45\textwidth}
\includegraphics[width=\textwidth]{7dRmpoint5StressTensor.png}
\caption{$R_-/R_+=.5$}
\end{subfigure}
\begin{subfigure}[b]{.45\textwidth}
\includegraphics[width=\textwidth]{7dRm1point0StressTensor.png}
\caption{$R_-/R_+=1.0$}
\end{subfigure}
\begin{subfigure}[b]{.45\textwidth}
\includegraphics[width=\textwidth]{7denergydensities.png}
\caption{}
\label{fig:7denergydensities}
\end{subfigure}
\caption{(a)-(c) are plots of the six dimensional $\frac{8\pi G_6}{3l^5}\langle T^{\mu}_{\;\;\nu}\rangle$ as a function of $R$ for different values of $R_-/R_+$. In the plots, the thick line is the energy density $T^{t}_{\;\;t}$, the dotted line is $T^{r}_{\;\;r}$, and the dashed line is $T^{\Omega}_{\;\;\Omega}$. In (d), we plot the energy densities for $R_-/R_+ = $0, .5, .9, and 1.0 (black, dotted line).}
\label{fig:7dstress}
\end{figure}
$\newline$

Next, the five dimensional stress tensor is given by
\begin{equation}
\langle T^{\mu}_{\;\;\nu} \rangle = \frac{5l^4}{16\pi G_6}\frac{1}{R^5}\text{diag}\biggl\{\frac{T_5(R)}{(1 + \frac{R_+}{R}) (1-\frac{R_-^2}{R^2})},-\frac{T_5(R)}{(1 + \frac{R_+}{R}) (1-\frac{R_-^2}{R^2})}-3S_5(R),S_5(R),S_5(R),S_5(R)\biggr\}
\end{equation}
This is traceless, as it should be, because there is no conformal anomaly in odd dimensions. In fig. \ref{fig:6dstress} we plot this for some choices of $R_-$. Here we note some differences from the four dimensional result. The first is that  the energy density starts negative near the horizon but becomes positive away from the horizon for smaller ratios of $R_-/R_+$. In higher dimensions, it seems as though the ``jammed" plasma is more easily localized away from the black hole. This is confirmed by the pressure becoming positive in this same region. As in the four dimensional case, the energy density first increases then decreases as $R_-/R_+ \to 1$. 
$\newline$

The six dimensional stress tensor is very messy, and so we will leave the full expression to the appendix. Here, we just note that the trace,
\begin{equation}
\langle T^{\mu}_{\mu}\rangle = \frac{3l^5}{8\pi G_6}\frac{9 R_-^6R_+^6 \left(280 R^6-320 R^3 \left(R_-^3+R_+^3\right)+271 R_-^3R_+^3\right)}{200 R^{24}}
\end{equation}
exactly matches the conformal anomaly, $a^{(6)}$, in 6 dimensions. In fig. \ref{fig:7dstress}, we plot this stress tensor for different values of $R_-/R_+$ as before. Here, we see a new phenomenon. Near the black hole, the energy density is positive, but becomes negative away from the black hole, and then becomes positive again. This reinforces the idea that in higher dimensions there is a stronger tendency for the CFT to localize away from the black hole as $T_{BH}\to T_{\infty}$.

It is worth pointing out that in all dimensions, the behavior far from the black hole matches the corresponding Tangerlinhi behavior with a $R^{-(d+1)}$ fall-off. This rapid fall off gives a strong indication that the CFT corresponds to an Unruh or Boulware state. Furthermore, the dimension dependence of this fall-off suggests that the black hole affects the CFT closer to the horizon in higher dimensions. As we will show below, from work done on Wilson loops in holography \cite{Hirata:2006jx}, there is a ``confinement" scale for the $T_\infty$ plasma that tends to decrease in size in higher dimensions. These may conspire to explain the dimension dependence of localization seen in the energy densities and in the entanglement entropies below.


\section{Entanglement Entropies of Droplets}
In this section we seek to clarify some of the results of the previous section. In particular, we learned from the boundary stress tensor that there is a region near to the black hole horizon where the CFT energy density becomes negative. This state prevents heat flow between the black hole at temperature $T_{BH}$ and asymptotic plasma at temperature $T_{\infty}$. In the ``jammed" phase, then, there should be very little correlation between degrees of freedom near the horizon and degrees of freedom in the asymptotic plasma. 

One measure of these correlations is the entanglement entropy of a spatial subregion in the CFT. To define the entanglement entropy of a spatial subregion, one first takes a time slice of the field theory manifold on which to define a Hilbert space. On this time slice, one then chooses a spatial subregion which we will call $\mathcal{A}$ and divides the Hilbert space into two subspaces, one that contains degrees of freedom purely within $\mathcal{A}$ another that contains only degrees of freedom in $\bar{\mathcal{A}}$, $\mathcal{H} = \mathcal{H}_{\mathcal{A}}\otimes\mathcal{H}_{\bar{\mathcal{A}}}$. Given a state on the full Hilbert space defined by a density matrix $\rho$, one can define a reduced density matrix, $\rho_A$, describing only $\mathcal{H}_\mathcal{A}$ by tracing out the degrees of freedom in $\mathcal{\bar{A}}$. The entanglement entropy is then given by the von Neumann entropy of this density matrix,\footnote{As mentioned earlier, this includes entanglement with a purifying system if $\rho$ is not pure.}
\begin{equation}
S_A = -Tr\rho_A\log\rho_A.
\end{equation}
One may check that if our Hilbert space defines a single entangled pair, the entanglement entropy of a region containing just one member of the pair is 2 but a region containing both pairs is 0. Thus, at least to first order, the entanglement entropy quantifies how correlations are shared across the boundary of the spatial region $\partial \mathcal{A}$. 

For free field theories, one can occasionally calculate the entanglement entropy. These calculations often require the computation of so-called Renyi entropies which come from the analytic continuation of path integrals on Riemann surfaces  \cite{Calabrese:2004eu, Klebanov:2011uf}.  However, for interacting field theories, such calculations become more burdensome, especially for theories without large numbers of symmetries. In the AdS/CFT correspondence, degrees of freedom on the boundary are often ambiguous and such calculations on the field theory side are prohibitive. Fortunately, for field theories on static spacetimes, Ryu and Takayanagi gave a procedure to calculate these entanglement entropies by solving for surfaces in the bulk \cite{Ryu:2006bv, Ryu:2006ef}. Given a spatial region of the field theory $\mathcal{A}$ with boundary $\partial \mathcal{A}$, one solves for a codimension-two minimal surface which starts on $\partial \mathcal{A}$ and extends into the bulk.\footnote{For our droplet solutions, the horizon stretching into the bulk acts as a barrier that minimal surfaces may not cross \cite{Engelhardt:2013tra}.} The surface which corresponds to the entanglement entropy of $\mathcal{A}$ is that surface which minimizes the area. Occasionally, there may be a ``phase transition" between two different types of surfaces which  minimize the area at different values of some parameter, for instance the width of the chosen spatial region \cite{Hirata:2006jx}. 

Note that due to the short range correlations of quantum field theories, entanglement entropies are strictly divergent. Defining a UV cutoff in the field theory, $\epsilon$, the leading order divergence is proportional to Area($\mathcal{A}$)/$\epsilon^{d-2}$ \cite{Srednicki:1993im}. For the holographic entanglement entropy, the field theory cutoff corresponds, on the bulk side, to evaluating the area of the minimal surface up to a fixed $z=\epsilon$ slice. In the following, we wish to analyze correlations between the jammed and asymptotic plasma and so we calculate the entanglement entropy of ball shaped regions on the boundary as a function of the radius of the ball.\footnote{There are different invariant measures of radii on the boundary including the distance from the horizon and the coefficient of $d\Omega^2_{d-2}$ in $\tilde{g}_{\mu\nu}$. We choose the latter as it is finite, even as $R_-\to R_+$} We present both the UV divergences of this quantity as well as the regularized entanglement entropies where these divergences are subtracted. 

\subsection{AdS C-metric}
The AdS C-metric is an analytic solution to the Einstein equations in four bulk dimensions with negative cosmological constant \cite{Hubeny:2009kz}. For different regions of parameter space, this metric has both droplets and funnel solutions with hyperbolic black holes on the boundary. For one particular choice of parameters, however, the C-metric gives a droplet with an asymptotically flat black hole on the boundary. The metric for this droplet is 
\begin{equation}
\begin{split}
\label{eq:cmetric}
ds^2 &= \frac{l^2}{(x-y)^2}\left(-F(y)dt^2+\frac{dy^2}{F(y)}+\frac{dx^2}{G(x)} +G(x)d\phi^2\right)\\
\text{with } &F(y) = y^2+2\mu y^3,\quad G(x)=1-x^2-2\mu x^3 = 1-F(x)
\end{split}
\end{equation}
This spacetime is asymptotically locally AdS with conformal boundary at $x=y$. For $x-y\geq 0$ and $-\frac{1}{2\mu} \leq y \leq 0$, the spacetime has black hole on the boundary that extends into the bulk and touches the axis of rotation symmetry ($G(x_0)=0$). This black hole has a temperature $T = 1/4\pi\mu$. For $0\leq y\leq x_0$, we have a similar spacetime with no black hole on the boundary. For a given $\mu$, both spacetimes have an equal conical deficit $\Delta\phi = \frac{4\pi}{|G'(x_0)|}$. They both also have a zero temperature horizon in the bulk at $y=0$. This spacetime is only well defined for $\mu \geq \frac{1}{3\sqrt{3}}$, below which $G(x)$ is not positive semi-definite.

Conformally rescaling by $1/x^2$ and defining $\xi = -1/x$, we find a boundary metric
\begin{equation}
ds_{\partial}^2 = -(1-\frac{2\mu}{\chi})dt^2 + \frac{d\chi^2}{(1-\frac{2\mu}{\chi})G(-1/\chi)}+\chi^2G(-1/\chi)d\phi^2.
\end{equation}
As $\chi\to \infty$, $G(-1/\chi)\to 1$ and we see this metric describes flat 3 dimensional Minkowski space. Note that in this conformal frame, the horizon has area $A_{h} = \frac{8\pi\mu}{|G'(x_0)|}$. For completeness, we also note that for $\mu=0$, the C-metric gives Poincar{\'e}-AdS$_4$. Defining
\begin{equation}
r=\frac{\sqrt{G(x)}}{y},\quad z=(\frac{x}{y}-1)
\end{equation}
the metric (\ref{eq:cmetric}) becomes
\begin{equation}
ds^2 = \frac{l^2}{z^2}(-dt^2 + dz^2 + dr^2 +r^2d\phi^2).
\end{equation}
To find minimal surfaces, we first need to define a disc shaped region on the boundary. There are a couple choices of invariant radii, but the one that seems to make the most sense is the circumference radii $R=\frac{\sqrt{G(x)}}{x}$.\footnote{This radius only monotonically increases for $\mu\geq \frac{1}{2\sqrt{2}}$ which will be the range we will investigate.} Thus a choice of $R$ defines a choice $x_b=y_b$ on the boundary. Then we minimize the area functional
\begin{equation}
\frac{A}{\Delta \phi} = \int_{x_b+\epsilon_x}^{x_0} dx \frac{1}{(x-y(x))^2}\sqrt{\frac{G(x)}{F(y)}y'(x)^2+1}
\end{equation}
where $\epsilon_x$ is proportional to the UV cut-off in the field theory. Now, a defining feature of minimal surfaces is that they are normal to the conformal boundary. This tells us that 
\begin{equation} 
y'(x)|_{x=x_b} = -\frac{F(x_b)}{G(x_b)}\quad\text{and}\quad\epsilon = \frac{\epsilon_x}{x_bG(x_b)}
\end{equation}
where $z=\epsilon$ is the fixed field theory cutoff. We Taylor expand our curve $y(x)$ near $y=x_b$ and plug this into the area functional to see
\begin{equation}
\frac{A}{\Delta\phi}=\int_{x_b+\epsilon_x} \frac{dx}{(x-x_b)^2}(1+\frac{F(x_b)}{G(x_b)})^{-3/2} \to \frac{R_b}{\epsilon} + \text{finite}
\end{equation}
The divergence is the same for both the spacetime with a boundary black hole and without a boundary black hole, but the finite piece is different.

\begin{figure}[t]
\centering
\begin{subfigure}[b]{.45\textwidth}
\includegraphics[width=\textwidth]{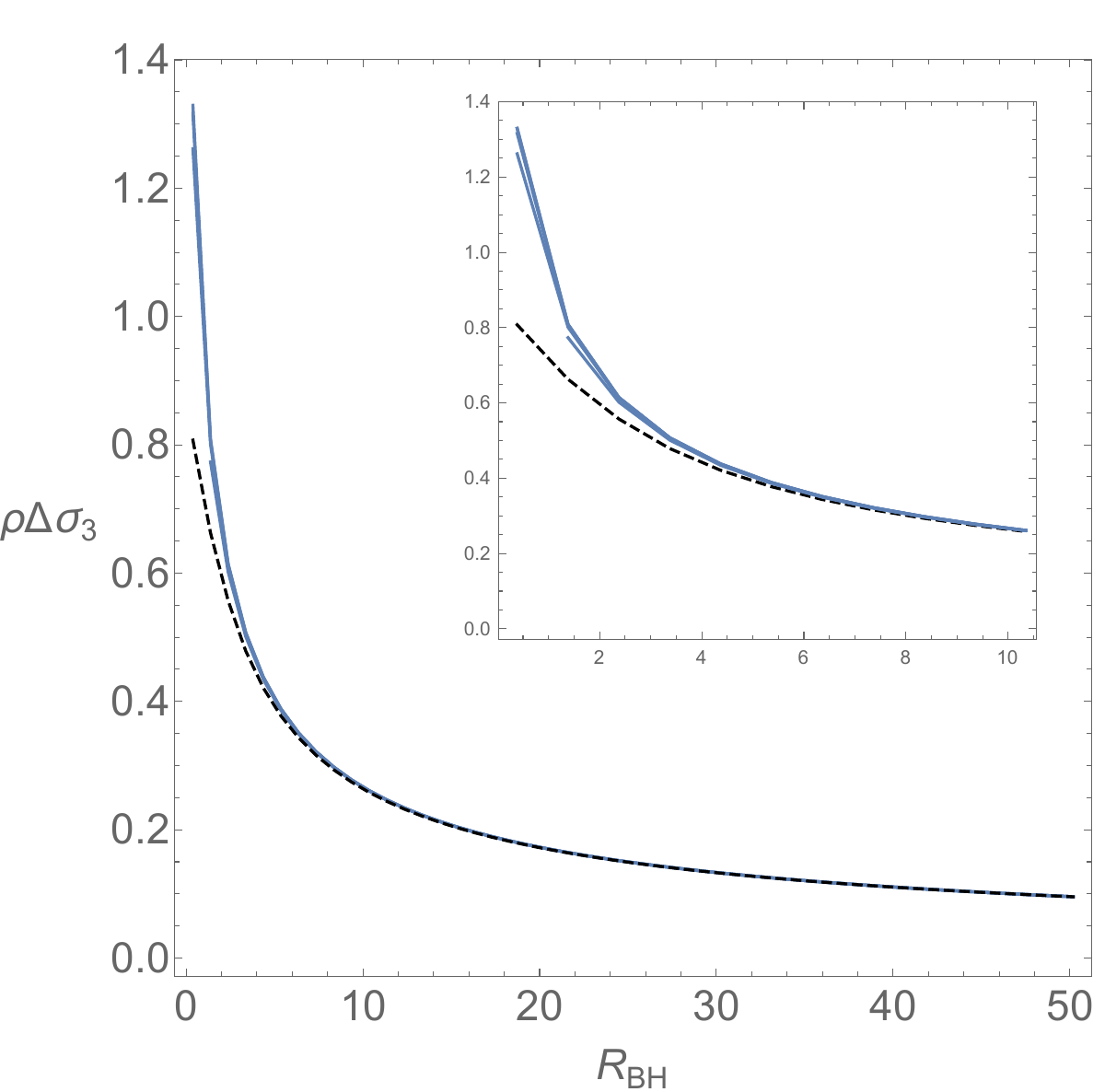}
\caption{}
\end{subfigure}
\begin{subfigure}[b]{.45\textwidth}
\includegraphics[width=\textwidth]{cmetricloglog.png}
\caption{}
\label{fig:3dhorizonlog}
\end{subfigure}
\caption{In (a), we plot $\rho\Delta\sigma_3$ for $\rho = 1$ (black, dotted line) $,2,3,10,20,100$ as a function of $R_{BH}$. In scaling by $\rho$, the area curves nearly perfectly overlap and the entanglement entropy has a universal behavior. Near the horizon (inset plot), there is some deviation, especially for $\rho = 1$ (black dotted line) but the discrepancy disappears for $\rho$ larger than $\sim 2$ ($\rho$ increases from bottom to top). In (b), we show a log-log plot of the horizon entanglement as a function of $T_{BH}=(4\pi \mu)^{-1}$. The line that we plotted shows this grows as $T_{BH}^{2/3}$.}
\label{fig:3dresults}
\end{figure}

To understand how the black hole affects correlations in the field theory, we want to compare entanglement entropies for equivalent size regions in both spacetimes. A convenient way to do so is to fix the radius of the disc $R_b$ and subtract the entanglement entropy for the field theory with no boundary black hole, $S_{NBH}$, from the entanglement entropy for the field theory with a boundary black hole.\footnote{In this paper, we will write everything in terms of the areas, leaving out the factor of $1/4G_{d+1}$ which would give the entropy. The no black hole spacetime is the $x>0$ region for a given $\mu$, not three dimensional Minkowski space.} This gives a finite value for the entanglement entropy
\begin{equation}
\Delta\sigma_3 \equiv \frac{A_{BH}(R_b)-A_{NBH}(R_b)}{2\pi}.
\end{equation}
This subtraction is standard in the literature, for instance in understanding thermal correlations in a strongly coupled field theory, one subtracts the entropy of pure AdS from the entropy of AdS-Schwarzschild for identical boundary regions. Conveniently, this method of comparison also gets rid of the divergence in the entanglement entropies. Notably, the cancellation of all divergences is particular to $d=3$. We will show below that in higher dimensions, it is only the leading order divergence which is cancelled. 

In fig. \ref{fig:3dresults}, we plot the entanglement entropy for a fixed ratio $\rho=R_b/R_{BH}$ and vary the black hole radius $R_{BH}=2\mu$. For $\mu > \sim1.5$, or for $\rho> \sim1.5$, the different entanglement entropies as a function of $\mu$ vary only by the ratio of their radii $\rho$, 
\begin{equation}
A(\rho_1,\mu)=\frac{\rho_2}{\rho_1}A(\rho_2,\mu)
\end{equation}
In particular, one can find the entanglement entropy as a function of $R$ if one knows how the entropy of the horizon scales as a function of $\mu$. The dependence on $\rho$ agrees with the picture of a jammed phase where at larger $R_b$, there are fewer correlations between degrees of freedom at large radius and degrees of freedom near the black hole. Interestingly, as $T_{BH} \to T_{\infty}$, the difference in entanglement entropy at the horizon also goes to zero. We show a log-log plot of this quantity in fig. \ref{fig:3dhorizonlog} as a function of the black hole temperature. In this plot, we see that as we approach extremality, the correlations between degrees of freedom inside the horizon and those outside go to zero. Because of the $\rho$ dependence of the entanglement entropy, the vanishing of the horizon entanglement entropy tells us that $\Delta\sigma_3$ vanishes everywhere in the extremal limit. One might expect this if the extremal limit corresponds to the field theory approaching a zero-temperature Hartle-Hawking. 

\subsection{Numerical Solutions}
We now seek to answer whether the same behavior occurs in higher dimensions. Like the ``charge" $\mu$ in the C-metric, we will vary $R_-$ in (\ref{eq:Deltadofr}) to change $T_{BH}$. To find the minimal surfaces, we minimize the area functional 
\begin{equation}
\frac{A_d}{\Omega_{d-2}} = \int^{x_{max}}_0 \frac{dx}{1-x^2}\left(\frac{x^2g(x)S}{(1-x^2)^2}\right)^{d/2-1}\sqrt{\frac{4A}{(1-r(x)^2)^2}r'(x)^2 + \frac{4B}{g(x)}+\frac{2xr(x)F}{1-r(x)^2}r'(x)}
\end{equation}
where $f(r)$ and $g(x)$ were defined in (\ref{eq:deffandg}). Given a UV cutoff $z=\epsilon$, the cutoff in the coordinate $x$ is $x_{max} = \sqrt{1-\epsilon/R_b}$ where $R_b$ is the radius of the ball whose entanglement we are investigating. The integration limits are consistent with the change of variables for pure AdS. Note that $R_b$ is related to $r$ through the same coordinate definition (\ref{eq:boundaryRdefinition}). Using the expansions (\ref{eq:asymptotic4}), (\ref{eq:asymptotic5}), (\ref{eq:asymptotic6}) and the fact that the surfaces are normal to the boundary ($r'(x)\to 0$) the divergent terms in the entanglement are, 
\begin{figure}[t]
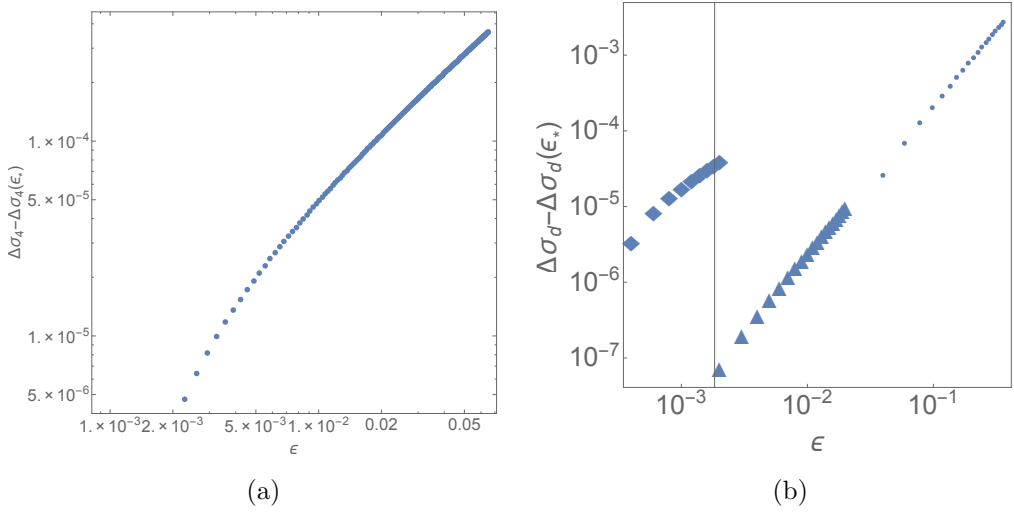

\centering
\begin{subfigure}[b]{.45\textwidth}
\includegraphics[width=\textwidth]{3dentropyerrors.png}
\caption{}
\end{subfigure}
\begin{subfigure}[b]{.45\textwidth}
\includegraphics[width=\textwidth]{entropyerrors.png}
\caption{}
\label{fig:entanglementerror}
\end{subfigure}
\caption{Above we show in a log-log plot the error in extracting $\Delta\sigma_d$ as a function of the cutoff, $\epsilon$. In (a), we do this for d=3 for $\mu=3$ and $R=R_+$. In (b), we do this for $R/R_+=3$ in $d=4$ ($\blacklozenge$), $d=5$ ($\blacktriangle$), and $d=6$ ($\bullet$). In all cases, we see a power-law convergence.}
\label{fig:NumericalExtraction}
\end{figure}
\begin{equation} 
\begin{split}
A_4/\Omega_{2} &\to \frac{1}{2}\frac{R_b^2}{\epsilon^2} - \frac{1}{2}(\alpha(r_b)-1)log(\epsilon)+\text{finite}\\
A_5/\Omega_{3} &\to \frac{1}{3}\frac{R_b^3}{\epsilon^3}-\frac{(8+3\beta(r_b))}{8}\frac{R_b}{\epsilon}+\text{finite}\\
A_6/\Omega_{4} &\to \frac{1}{4}\frac{R_b^4}{\epsilon^4}-\frac{(2\psi(r_b)+15)}{20}\frac{R_b^2}{\epsilon^2}-\frac{(15 + 20 \chi_r(r_b) - 12\psi(r_b) + 80 \chi_s(r_b))}{40}log(\epsilon)+\text{finite}
\end{split}
\end{equation}
At the black hole horizon, r=0, so that the above expansions are
\begin{equation} 
\begin{split}
A_4/\Omega_{2} &\to \frac{1}{2}\frac{R_{BH}^2}{\epsilon^2} +\frac{1}{2}(\frac{R_-}{R_+})log(\epsilon)+\text{finite}\\
A_5/\Omega_{3} &\to \frac{1}{3}\frac{R_{BH}^3}{\epsilon^3} -\frac{(2+9\frac{R_-^2}{R_+^2})}{8}\frac{R_{BH}}{\epsilon}+\text{finite}\\
A_6/\Omega_{4} &\to \frac{1}{4}\frac{R_{BH}^4}{\epsilon^4}-\frac{(5+18 \frac{R_-^3}{R_+^3})}{20}\frac{R_{BH}^2}{\epsilon^2}+ \frac{50 - 9 \frac{R_-^3}{R_+^3} (40 + 29 \frac{R_-^3}{R_+^3})}{200}log(\epsilon)+\text{finite}
\end{split}
\end{equation}
In particular, the leading order divergence is proportional to the area of the black hole horizon on the boundary, a similarity to the Bekenstein-Hawking entropy first noted in \cite{Emparan:2006ni, Solodukhin:2006xv}. 

One may also ask about the divergences as $R_b\to\infty$. We note that in this limit, $\alpha(r), \beta(r), \psi(r), \chi_r(r), \chi_s(r)$ all vanish and so
\begin{equation} 
\begin{split}
\label{eq:asymptoticlimit}
A_4/\Omega_{2} &\to \frac{1}{2}\frac{R_b^2}{\epsilon^2} + \frac{1}{2}\log(\epsilon)+\text{finite}\\
A_5/\Omega_{3} &\to \frac{1}{3}\frac{R_b^3}{\epsilon^3}-\frac{R_b}{\epsilon}+\text{finite}\\
A_6/\Omega_{4} &\to \frac{1}{4}\frac{R_b^4}{\epsilon^4}-\frac{3}{4}\frac{R_b^2}{\epsilon^2}-\frac{3}{8}log(\epsilon)+\text{finite}
\end{split}
\end{equation}
\begin{figure}[t]
\centering
\begin{subfigure}[b]{.45\textwidth}
\includegraphics[width=\textwidth]{4dentropyofr.png}
\caption{d=4}
\end{subfigure}
\begin{subfigure}[b]{.45\textwidth}
\includegraphics[width=\textwidth]{5dentropyofr.png}
\caption{d=5}
\end{subfigure}
\newline
\begin{subfigure}[b]{.45\textwidth}
\includegraphics[width=\textwidth]{6dentropyofr.png}
\caption{d=6}
\end{subfigure}
\begin{subfigure}[b]{.45\textwidth}
\includegraphics[width=\textwidth]{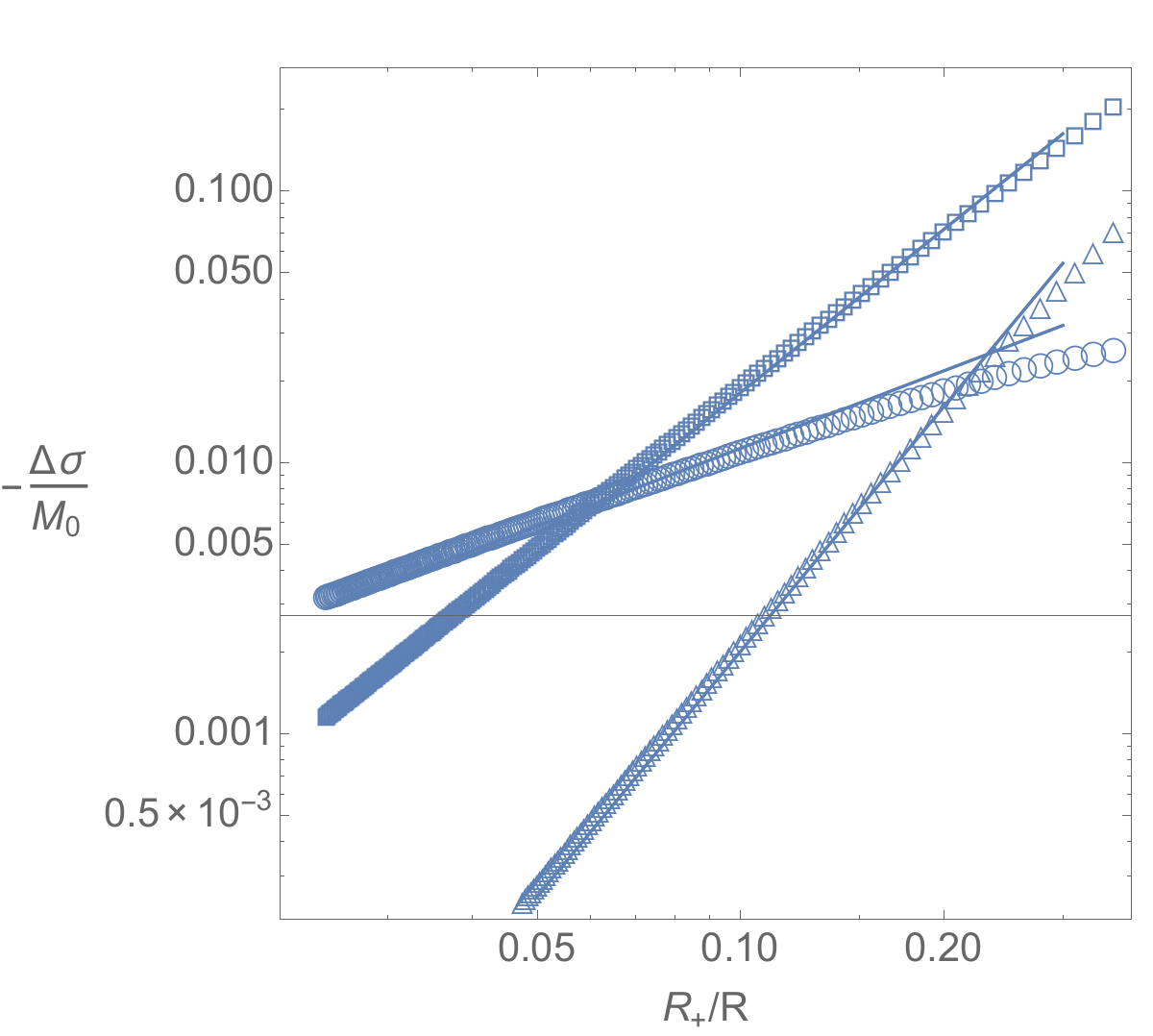}
\caption{}
\label{fig:loglogplot}
\end{subfigure}
\caption{In (a)-(c), we plot the finite piece of the entanglement entropies, defined in (\ref{eq:deltaA}), divided by the mass parameter $M_\kappa$ for a given $R_-$. As can be seen, far from the black hole, all entanglement entropies agree. In (a), the curves correspond to $R_-/R_+ =$ 0, .2, .4, .5, .8, 1.0 (top to bottom). In (b), $R_-/R_+=$ 0, .3, .5, .7, .8, 1.0 (bottom to top). Finally, (c) is the six dimensional case with $R_-/R_+ = $0, .4, .5, .6, .7, .8, .9, 1.0 (bottom to top). In all cases, the result for the extremal boundary black hole is in black and dashed. In (d), we show a log-log plot of the asymptotic part of the entanglement entropies as a function of $R_+/R$. These are linear ($\bigcirc$), quadratic ($\square$), and cubic ($\triangle$)for $d=4,5,6$ respectively.}
\label{fig:entropiesofr}
\end{figure}
\begin{figure}[t]
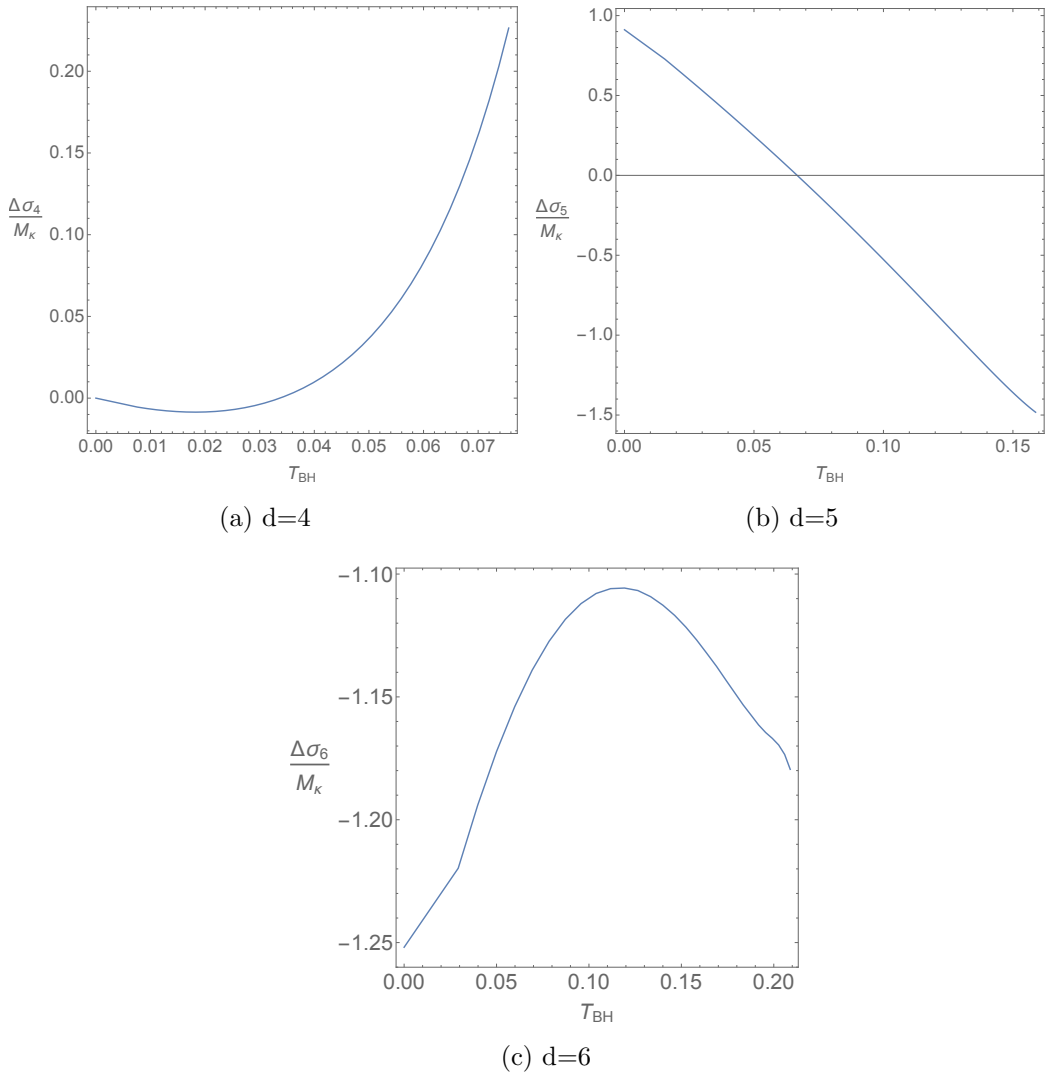

\centering
\begin{subfigure}[b]{.45\textwidth}
\includegraphics[width=\textwidth]{4dhorizonentropy.png}
\caption{d=4}
\end{subfigure}
\begin{subfigure}[b]{.45\textwidth}
\includegraphics[width=\textwidth]{5dhorizonentropy.png}
\caption{d=5}
\end{subfigure}
\newline
\begin{subfigure}[b]{.45\textwidth}
\includegraphics[width=\textwidth]{6dhorizonentropy.png}
\caption{d=6}
\end{subfigure}
\caption{In (a)-(c), we plot the finite piece of the entanglement entropies for the horizon as a function of $T_{BH}$. }
\label{fig:horizon}
\end{figure}
$\newline$
We want to subtract the entanglement entropies for balls of radius $R_b$ in pure AdS$_{d+1}$.\footnote{In four dimensions, this was calculated for instance in \cite{Hirata:2006jx} but the higher dimensional results are new} Here, we minimize
\begin{equation}
A_d = \Omega_{d-2}\int_{\epsilon}^{R_b} dz \frac{R(z)^{d-2}}{z^{d-1}}\sqrt{R'(z)^2+1}.
\end{equation}
In all dimensions, the minimal surface is $R(z) = \sqrt{R_b^2-z^2}$. Plugging this into the above expression gives
\begin{equation}
\begin{split}
\label{eq:pureAdS}
A_4/\Omega_{2} &= \frac{1}{2}\frac{R_b^2}{\epsilon^2} +\frac{1}{2}log(\epsilon) -\frac{1}{4}(1+2log(2))\\
A_5/\Omega_{3} &= \frac{1}{3}\frac{R_b^3}{\epsilon^3} - \frac{R_b}{\epsilon} +\frac{2}{3}\\
A_6/\Omega_{4} &= \frac{1}{4}\frac{R_b^4}{\epsilon^4} - \frac{3}{4}\frac{R_b^2}{\epsilon^2} - \frac{3}{8}log(\epsilon)+\frac{3}{32}(3+4log(2))
\end{split}
\end{equation}
As can be seen, the pure AdS divergences match the asymptotic limit of the droplet divergences (\ref{eq:asymptoticlimit}), as would be expected since the boundary spacetime is asymptotically flat. 

Now, as we did for the C-metric, we subtract the area of the surface in the no black hole background from the area of the surface in the black hole background. This gives
\begin{equation}
\begin{split}
\label{eq:deltaA}
\Delta A_4/\Omega_{2} &= -\frac{1}{2}\alpha(r_b)log(\epsilon) +\Delta \sigma_4\\
\Delta A_5/\Omega_{3} &= -\frac{3\beta(r_b)}{8}\frac{R_b}{\epsilon} +\Delta \sigma_5\\
\Delta A_6/\Omega_{4} &= -\frac{\psi(r_b)}{10}\frac{R_b^2}{\epsilon^2}-(2\chi_s(r_b)+\frac{1}{2}\chi_r(r_b)-\frac{3}{10}\psi(r_b))\log(\epsilon)+\Delta \sigma_6
\end{split}
\end{equation}
where $\sigma_d$ is the finite part of the d-dimensional area divided by $\Omega_{d-2}$. As can be seen, the leading divergence is gone but we still have to contend with subleading divergences. Our regularization procedure will be to fix a value of $\epsilon$ and then subtract these divergences from the area integral of our solutions. As we show in fig. \ref{fig:entanglementerror}, we find that $\Delta\sigma_d$ has power law convergence up to some minimum $\epsilon_*$. Below this value, the numerics become unstable. The plots of $\Delta\sigma_d$ come from $\epsilon$ slightly above this minimum.

In fig. \ref{fig:entropiesofr}, we plot our results for $\sigma_d/M_\kappa$ as a function of $R/R_+$ for various values of $R_-$ in $d=4,5,6$ ($M_\kappa$ is the mass appearing in $\Delta_d$ (\ref{eq:Deltadofr})). While the three plots look very similar, there are actually some important differences, especially between the $d=4$ and $d>4$ cases. We first note the similarities. In each case, the finite piece of the entanglement entropy is negative far from the black hole. Furthermore, once scaled by the mass, $M_\kappa$, the entanglement entropies agree for $R_b\gg R_+$. Note that we have already used conformal symmetry to set $R_+=1$ and so this procedure is analogous to the scaling by $\rho$ in the C-metric example. As we saw there, the entanglement entropy becomes universal (i.e. independent of $R_-$) in this limit. This shows that far from the horizon, the leading order fall-off in the metric determines the entanglement entropy. This is demonstrated in fig. \ref{fig:loglogplot} where we show that in this region, the entanglement entropies fall off as $\left(R_+/R\right)^{d-3}$, exactly following $\Delta_{d}(R)$ at large R. Importantly, this is not the case near the horizon, where interactions between the CFT and Hawking radiation have a strong influence on the entanglement entropy and as we show below, prevent heat flow from the black hole to spatial infinity.

In particular, near the horizon, we see a minimum (also a maximum in six dimensions) that becomes more exaggerated as $T_{BH}\to T_\infty$. This dip defines a radius, $R_{min}$, where entanglement between degrees of freedom at $R<R_{min}$ and $R>R_{min}$ is minimized. One explanation for these dips is that the jammed plasma is becoming more localized near these critical radii, $R_{min}$, more effectively blocking the flux of Hawking radiation to infinity. As discussed above, if a Bell pair is separated across the boundary of our entangling region, the entanglement entropy is two, whereas if both members of the pair are within the entangling region, the entropy is zero. If there is localization in the plasma, then there should be a region where correlations are dense. At $R = R_{min}$, the internal correlations of the localized plasma are all within our entangling region and so the entanglement entropy is at a local minimum.

\subsection{Localization in pure AdS}

\begin{figure}[t!]
\centering
\begin{subfigure}[b]{.32\textwidth}
\includegraphics[width=\textwidth]{4dconnectedvdisconnected.png}
\caption{d=4}
\end{subfigure}
\begin{subfigure}[b]{.32\textwidth}
\includegraphics[width=\textwidth]{5dconnectedvdisconnected.png}
\caption{d=5}
\end{subfigure}
\begin{subfigure}[b]{.32\textwidth}
\includegraphics[width=\textwidth]{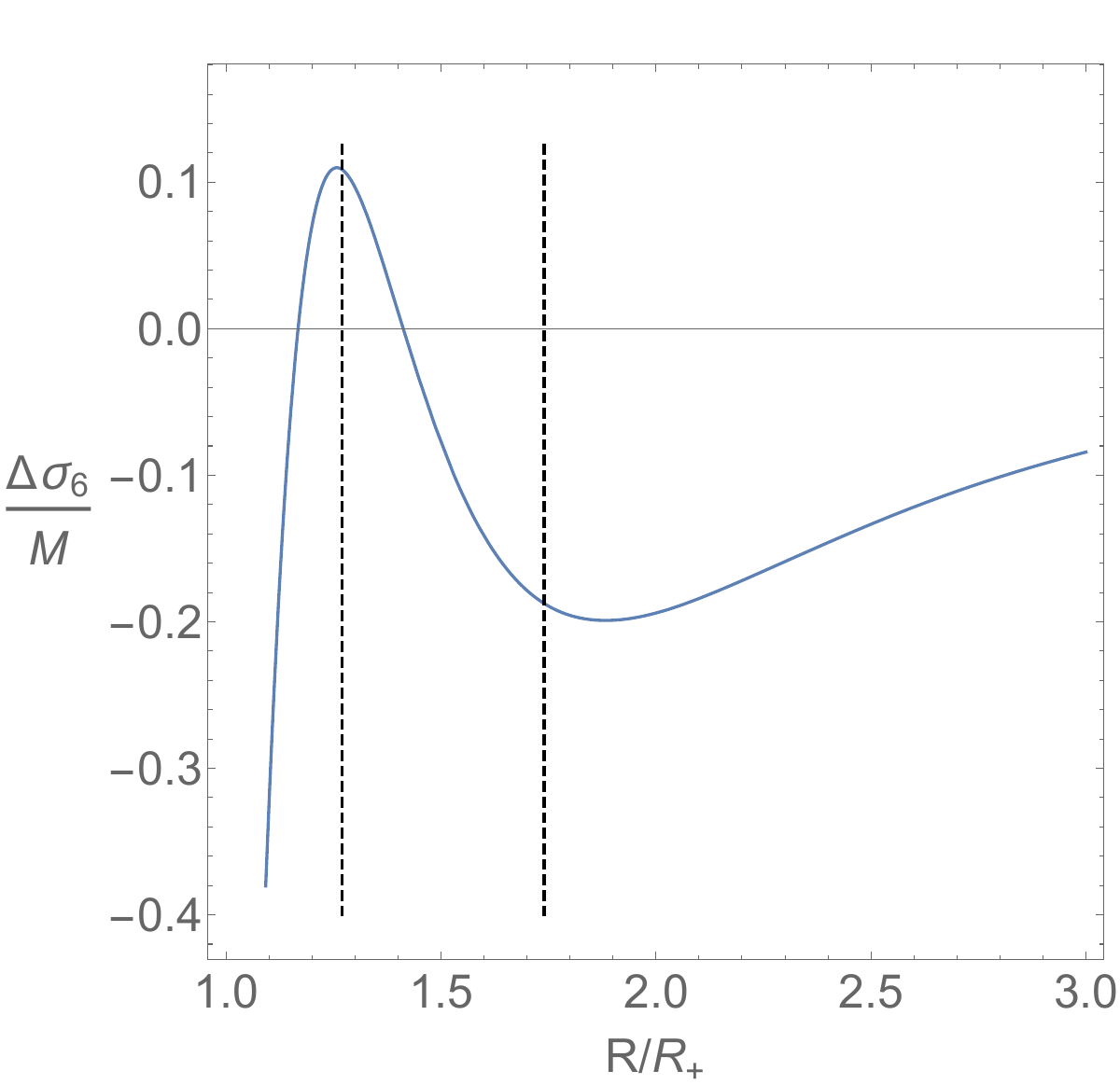}
\caption{d=6}
\end{subfigure}
\newline
\begin{subfigure}[b]{.32\textwidth}
\includegraphics[width=\textwidth]{4dentanglementmin.png}
\caption{d=4}
\end{subfigure}
\begin{subfigure}[b]{.32\textwidth}
\includegraphics[width=\textwidth]{5dentanglementmin.png}
\caption{d=5}
\end{subfigure}
\begin{subfigure}[b]{.32\textwidth}
\includegraphics[width=\textwidth]{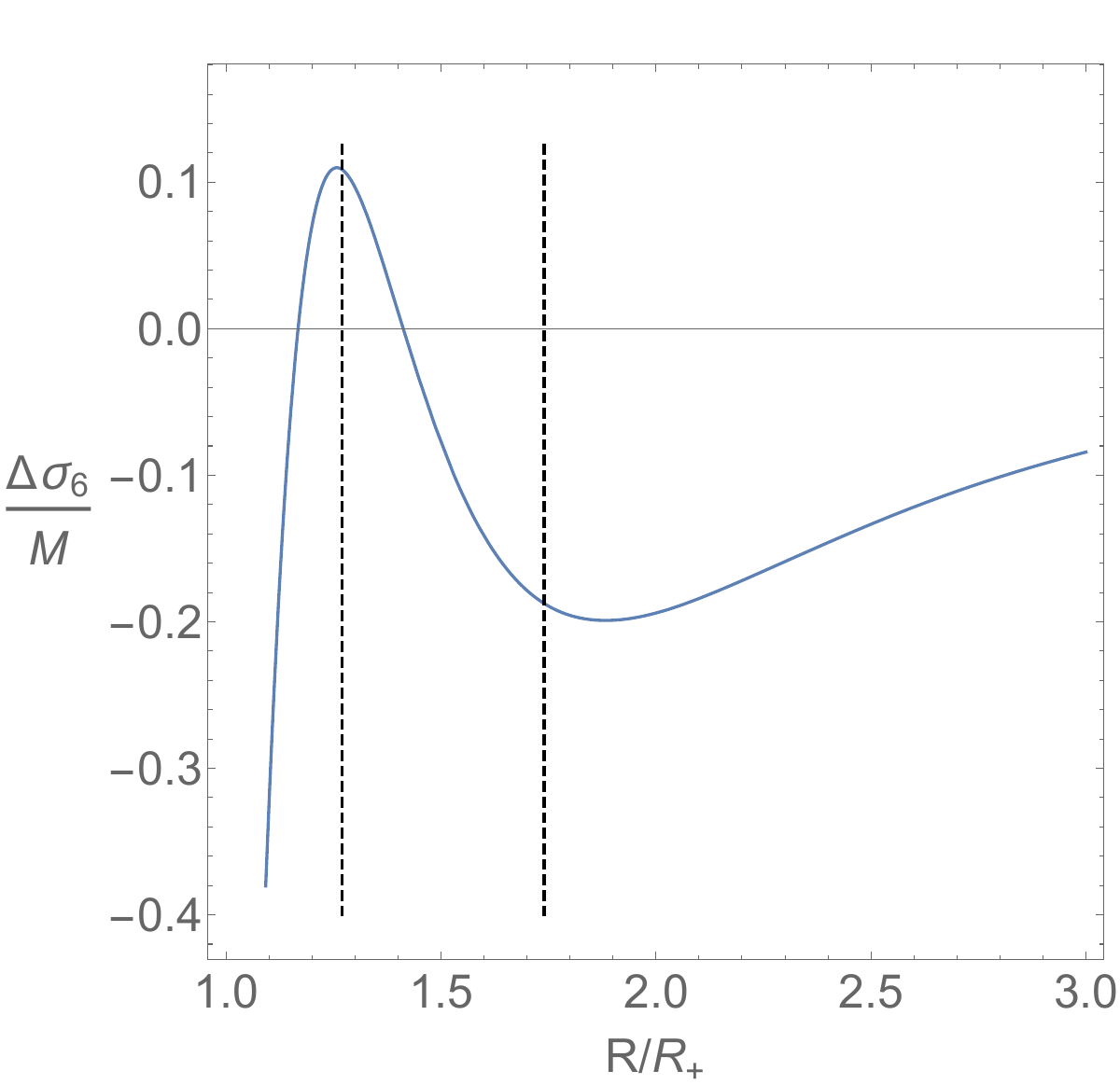}
\caption{d=6}
\end{subfigure}
\caption{In (a)-(c), we plot the area of the minimal surfaces for boundary annuli as a function of the ratio of outer and inner radii $\rho=\frac{R_2}{R_1}$. The dotted line is the area for the ``connected" surface while the filled line is the area for two ``disconnected" balls. At some critical $\rho^{(d)}_*$ the minimal area surface changes, representing a confining phase transition. In (d)-(f), we show the entanglement entropy of balls in the extremal black hole background $\Delta\sigma_d$ and mark $\rho^{(d)}_*$. In $d=6$, the first line marks radius where $\Delta\sigma_6$ is maximized and the second line is $\rho^{(6)}_*$ times this radius.}
\label{fig:connecteddisconnected}
\end{figure}

To better understand this picture of localization in the $T_{BH}$ plasma state, we seek to quantify a length scale for confinement in the CFT vacuum (corresponding to the $T_\infty=0$ state). We extend the work of \cite{Hirata:2006jx} to find a confining phase transition in $d\geq 4$ Poincar{\'e}-AdS. It is worth noting that this spacetime is not compact, so this is not the usual holographic picture of confinement in a CFT \cite{Witten:1998zw}. However, it was shown in \cite{Gross:1998gk} that using holography one can relate a glueball mass in four-dimensional QCD to a phase transition in the two-point correlator of Wilson loops in $\mathbb{R}^4$ at finite temperature. The correlator can be obtained from finding the minimal surface connecting two loops separated by a distance $L$. For $L$ larger than some critical $L_c$, the minimal surface becomes disconnected and the correlator vanishes, an indication of confinement.\footnote{Nominally, this implies the glueball mass is infinite. In fact, one expects that when the cross section of the minimal surface in this geometry is on the order of the string-scale, the supergravity approximation breaks down and instead the correlator is dominated by supergraviton exchange between the Wilson loops. The extension of this Gross-Ooguri phase transition to holographic geometries dual to confining CFTs was performed in \cite{Klebanov:2007ws}} This transition can also be related to monopole condensation in four dimensions. For monopoles separated by a distance $L>L_c$, the phase transition in the minimal surface shows that the potential between the monopoles goes to a constant, effectively screening the monopoles from each other so that the force between them vanishes. It seems possible, then, that a similar phase transition in the CFT on a black hole background may act to prevent heat exchange between the black hole and an asymptotic plasma.

To find the critical length scale for our droplets, we start with Poincar{\'e}-AdS$_{d+1}$ and study the entanglement entropy of annuli on the boundary with inner radius $R_1$ and outer radius $R_2$. As we vary the ratio $\rho=(\frac{R_1}{R_2})$, there is again phase transition between ``connected" and ``disconnected" surfaces. For small $\rho$, the surfaces which minimize the area are ``connected". Using a conformal transformation to take the plane to a cylinder, this surface can be understood as the string world sheet connecting two Wilson loops analogous to the construction in \cite{Gross:1998gk}.\footnote{One important feature of the construction in \cite{Gross:1998gk} was that the Euclidean supergravity solution had an $S^1$ from the Euclidean time circle. In our conformal map, the $S^1$ of the cylinder is analogous to the thermal circle.} At some critical $\rho^{(d)}_*$, the minimal area surface for the annuli is instead two disconnected surfaces, each of which corresponds to the ball surfaces discussed above. In the frame of the cylinder, this tells us that the string worldsheet no longer stretches between the two Wilson loops. As before, the area and therefore the two point correlator does not scale with $\rho$ and so this is understood to represent a type of confinement. The critical ratio $\rho_*$ is then understood to correspond to a size for the ``glueballs" of the CFT above which heat transport is screened.\footnote{As discussed above, the real length scale actually depends on the map from the plane to the cylinder.} 

As seen in fig. \ref{fig:connecteddisconnected}, in pure AdS, the phase transition for the $d=4$ plasma occurs at $\rho^{(4)}_*\equiv(R_2/R_1)_*\sim1.844$. As the boundary black hole nears extremality, we also see that a minimum in the entanglement entropy appears at $R/R_+ \sim\rho_*$. In $d=5$, we find that $\rho^{(5)}_* \sim 1.53$.  Again, as $T_{BH}\to 0$, the minimum in the entanglement entropy occurs at $R/R_+ \sim \rho^{(5)}_*$. We conjecture that this coincidence occurs because the boundary black hole breaks conformal symmetry and in our conformal frame, the event horizon sets a fundamental size for the jammed plasma. This lends some credence to the idea that the jamming occurs because the plasma cannot fit inside the black hole. In the confined phase, where the minimum radius is given by $R_+$, the jammed plasma can extend only to $\rho^{(d)}_*$. The ``glueball" of the plasma is highly entangled with the degrees of freedom behind the horizon. Once the radius that defines the boundary region is larger than $\rho^{(d)}_*$, there are no longer correlations beween the degrees of freedom behind the horizon and the glueball, hence the minimum in the entanglement. The entanglement then grows again because of correlations between degrees of freedom away from the black hole matching onto the Tangerlinhi behavior.

In the $d=6$ entanglement entropy, the behavior is slightly different and we now see a maximum and a minimum. From the stress tensor (fig. \ref{fig:7denergydensities}), we saw that the region of negative energy density has moved away from the horizon, unlike the lower dimensional cases. This suggests that it in this dimension, it may not be accurate to draw conclusions about localization scales in terms of $R_+$ for the entanglement entropy.\footnote{In even dimensions, there is some ambiguity in the regularized entanglement entropy because of the log divergence. While we use the same regularization scheme for each value of $R_-$, this ambiguity may still play a role. We thank Don Marolf for emphasizing this point to us.} In fact, the maximum entanglement occurs at $R_{max}/R_+\sim 1.25$. Multiplying this radius by $\rho_*^{(6)}\sim 1.346$ gives us approximately the position of the dip in entanglement. In this dimension, it is clear that there is a complicated relationship between the entanglement outside of the horizon and the jammed plasma. The positive energy density near the horizon suggests possible new degrees of freedom which are not highly entangled with the black hole causing the global minimum of entanglement to occur at the horizon in this dimension.\footnote{As the energy density is also positive far from the black hole, these degrees of freedom my behave very similarly to the plasma far from the black hole. Note that by dividing the Hilbert space into two spatial regions, we can't distinguish which degrees of freedom are correlated.} As we increase the radius of our entangling surface, the entanglement increases as the surface divides these near horizon degrees of freedom into two regions. This increase continues until $R_{max}$ where the glueball is located. The entanglement then behaves as the $d=4,5$ cases, decreasing until eventually reaching a local minimum where the glueball is entirely within the entangling surface and then increasing. 

Interestingly, in $d>4$, there are no local minima in the entanglement entropy for sufficiently large temperatures. Here it seems that the Hawking radiation prevents the plasma from confining.\footnote{In \cite{Kim:2001td}, it was shown that the critical $\rho_*^{(d)}$ increases with temperature.} In fact, the equality between the location of this dip measured in terms of $R_+$ and $\rho^{(d)}_*$ is particular to the extremal black hole where the Hawking radiation is at zero temperature and no longer excites the plasma. Here, the jammed CFT should behave similarly to the vacuum state and share the same confining scale. For higher temperature black holes, the plasma should be excited and its pressure should increase. At the same time, there is a stronger gravitational attraction which makes the plasma want to decrease in size. It appears that in four dimensions, the Hawking radiation wins out and the dip in the entanglement entropy moves to larger radii. In $d > 4$, the gravitational attraction wins and the dips move toward the horizon. It is worth noting that the stress tensors (figs. \ref{fig:5denergydensities}, \ref{fig:6denergydensities}, \ref{fig:7denergydensities}) confirm this behavior as all energy densities have peaks that mirror the temperature dependence of the entanglement entropy. 

Now we note some important differences. In $d=3$, we find $\Delta\sigma_3>0$ for all R. However, for $d>3$, $\Delta\sigma_d$ is negative except near the horizon. This negative $\Delta\sigma_d$ can be explained because the vacuum dual to Poincar{\'e} AdS is in a pure state and very highly entangled \cite{Casini:2011kv, Czech:2012be}.\footnote{It was shown in \cite{Casini:2011kv} that the density matrix for ball shaped regions in Poincar{\'e}-AdS can be conformally mapped to a thermal density matrix which is known to have maximal entanglement.}  It is reasonable that far from the black hole, the jammed CFT has less correlations than the vacuum. The new behavior near the horizon can be explained by the breaking of conformal invariance. The vacuum dual to Poincare{\'e}-AdS is scale invariant and on a fixed time slice, any division of the boundary into two pieces will give the same value for the finite piece of the entanglement entropy (\ref{eq:pureAdS}). The droplet spacetime, however, is in general not Ricci flat and breaks this scale invariance. It is worth noting that, in $d=5,6$, when the boundary is Ricci flat ($R_-=0$), the entanglement entropy is monotonic, similar to the three dimensional case, although this is not the case in $d=4$. Near the black hole, the entanglement becomes positive because the CFT is very localized and has denser correlations than the vacuum. As we increase $R$ away from the horizon, the entanglement decreases until $R/R_+=\rho^{(d)}_*$ where the glueball of the jammed CFT is completely within our boundary region. From here, the entanglement entropy matches the behavior of the Ricci flat case.

The three dimensional case is unique because neither the black hole nor the no black hole spacetime is scale invariant despite being Ricci flat.\footnote{A scale invariant, spherically symmetric spacetime would have the same value for the entanglement entropy for any choice of $R$.} While the $\mu=0$ limit of the C-metric gives Poincar{\'e}-AdS$_4$, it is more appropriate to subtract the no boundary black hole state because it has the same leading order divergence and the same conical deficit as the black hole spacetime for a given $\mu$. Furthermore, for large radii, $\Delta\sigma_3\to 0$. Here it is not clear whether one should expect the no black hole background, which corresponds to an excited state of the CFT, to have more or less entanglement than the black hole background.

The next difference is the behavior of the entanglement on the horizon (fig. \ref{fig:horizon}). It is intriguing that the entanglement is not monotonic in $T_{BH}$. In three and four dimensions, this quantity starts positive and goes to zero as we go to extremality. In four dimensions, the entanglement entropy actually becomes negative at some critical $T_{BH} \sim .02$ but then increases towards zero again. In five and six dimensions, the entanglement starts negative and increases. In five dimensions this continues all the way to $T_{BH}=0$ while the six dimensional case starts decreasing around $T_{BH} \sim .12$. In $d=5$, the entanglement entropy on the horizon is linear while in $d=4$ and $d=6$ it is roughly quadratic. In $d=3$, as seen in fig. \ref{fig:3dhorizonlog}, the entanglement entropy at the horizon increases as $T_{BH}^{2/3}$. We don't have a good understanding of this behavior. In particular, we note that for $d\geq 4$, $M, Q\to 1$ as $T_{BH}\to 0$, but the three dimensional case has $\mu \to \infty$ as $T_{BH}\to 0$ more similar to a Schwarzschild black hole where the extremal limit is $M\to \infty$. In fig. \ref{fig:embeddings}, we showed that in $d=4$, as $T_{BH}\to 0$, the horizon approaches the surface $R^2 + z^2 = 1$. In Poincar{\'e}-AdS, this is the minimal surface for a ball with radius $R=1$. Thus we expect that in the extremal limit, the difference in entanglement entropy should vanish as it does. This is not the case in higher dimensions where the surface cannot be isometrically embedded in this limit. It seems that the extremal limit of the horizon entanglement entropy is dimension dependent. One may hope to gain insight by doing similar calculations in other droplet spacetimes by fixing the boundary black hole temperature and varying $T_\infty$ where we can use a Schwarzschild black hole on the boundary and vary the bulk horizon temperature.\footnote{This is a conformal field theory, so that only the ratio $T_{BH}/T_{\infty}$ should matter. However, the black holes that we constructed have a maximum temperature because we set $R_+=1$.}

\section{Discussion}
In this paper we sought to clarify some features of jammed CFTs by investigating their holographic duals. To this end, we constructed three new classes of solutions to the Einstein equations with a negative cosmological constant that have boundaries conformal to the d-dimensional Reissner-Nordstr{\"o}m metric. 

For these solutions, we calculated the boundary stress tensor and compared this to both a theoretical form as well as to existing literature. The calculation of the boundary stress tensor introduced some new features that had not been seen before in the literature, including peaks in the energy densities, positive asymptotic energy densities, and magnitudes near the horizon that were not monotonic as a function of $T_{BH}$. This new behavior seems to indicate the presence of a CFT phase that becomes localized away from the horizon as $T_{BH}\to T_\infty$. 

We also calculated entanglement entropies of balls in these geometries as well as in the analytic example of the AdS C-metric. For these examples, many features of the boundary stress tensor were confirmed, including near horizon behavior indicative of a localized CFT. Furthermore, for $d\geq 4$, the entanglement entropies were smaller in the black hole background than the corresponding entropies in pure AdS far from the black hole showing very little correlation between a near horizon CFT and an asymptotic CFT.\footnote{Note that in work done on thermal states dual to the AdS-Schwarzschild black hole, differences are positive. This is because the thermal state, in addition to the vacuum entanglement, has new degrees of freedom from thermally excited entangled pairs.} Furthermore, by comparing to confining phase transitions in Poincar{\'e}-AdS$_{d+1}$, we find critical values in the entanglement entropy which match the proposed ``size" of glueballs in units of the horizon radius $R_+$. At the same time, many features of the entanglement entropy remain enigmatic. One particular example is the behavior of the horizon's entanglement entropy as a function of $R_-$. The different behavior between the $d=3$ case and $d\geq 4$ cases suggest that one direction worth pursuing is to choose a Schwarzschild boundary black hole and vary $T_\infty$ instead. Using the construction in \cite{Santos:2014yja}, one could better compare to the three dimensional case. In particular, one may understand why, in $d=5, d=6$, $\Delta\sigma_d$ does not go to zero as the boundary black hole approaches extremality.

\section*{Acknowledgments}
It is a pleasure to thank Netta Engelhardt, Gary Horowitz, and Don Marolf for helpful discussions on this work and valuable feedback on this manuscript. This work was supported in part by NSF grant PHY-1504541.


\appendix
\section{Stress Tensor Expansion}

In this appendix, we present the expansions used to calculate the boundary stress tensor. As discussed above, our spacetime is asymptotically locally AdS \cite{Fischetti:2012rd}. This suggests that in a finite neighborhood $U$ of the boundary $\partial\mathcal{M}$, we can define a ``Fefferman-Graham" coordinate, z, such that 
\begin{equation} 
z|_{x=1}=0, \;\;\hat{g}^{MN}\partial_{M}z\partial_{N}z = 1/l^2, \quad\text{where}\;\; \hat{g}=z^2g
\end{equation}
This coordinate allows us to construct Gaussian normal coordinates near the boundary such that the metric is given by
\begin{equation}
ds^2 = \frac{l^2}{z^2}(dz^2 + \gamma_{\mu\nu}(x,z)dx^\mu dx^\nu).
\end{equation} 
From the Einstein equations, one may show that the metric $\gamma_{\mu\nu}$ can be expanded near the boundary in even powers of $z$ up to order $z^d$,
\begin{equation}
\gamma_{\mu\nu}(x,z) = \gamma_{\mu\nu}^{(0)} + z^2\gamma_{\mu\nu}^{(2)}+...+z^d\gamma_{\mu\nu}^{(d)}+z^d\bar{\gamma}_{\mu\nu}^{d}\log z^2+...
\end{equation}
where the $\bar{\gamma}_{\mu\nu}^{(d)}$ term only appears for even $d$. Each term in the expansion up to $\gamma^{(d)}$ can be expressed in terms of geometric quantities determined from the boundary metric $\gamma_{\mu\nu}^{(0)}$. For example,
\begin{equation}
\begin{split}
\gamma_{\mu\nu}^{(2)} &= -\frac{1}{(d-2)}\left(\mathcal{R}_{\mu\nu} - \frac{1}{2(d-1)}\mathcal{R}\gamma_{\mu\nu}^{(0)}\right)\\
\end{split}
\end{equation}
 where $\mathcal{R}_{\mu\nu\rho\sigma}, \mathcal{R}_{\mu\nu},\mathcal{R}$ are the Riemann tensor, Ricci tensor, and Ricci scalar of the boundary metric respectively. Other expressions can be found in \cite{deHaro:2000vlm}.\footnote{These authors use a different convention for the Riemann tensor and some care must be taken to compare to our expressions.} Furthermore, one can express $\bar{\gamma}_{\mu\nu}^{(d)}$ in terms of these geometric quantities and the covariant derivatives associated with the boundary metric,
\begin{equation}
\begin{split}
\bar{\gamma}_{\mu\nu}^{(2)} &= 0\\
\bar{\gamma}_{\mu\nu}^{(4)} &= \frac{1}{8}R_{\mu\nu\rho\sigma}R^{\rho\sigma} - \frac{1}{48}\nabla_\mu\nabla_\nu\mathcal{R}+\frac{1}{16}\nabla^2R_{\mu\nu}-\frac{1}{24}\mathcal{R}\mathcal{R}_{\mu\nu}\\&\quad\quad\quad\quad + \left(-\frac{1}{96}\nabla^2\mathcal{R}+\frac{1}{96}\mathcal{R}^2-\frac{1}{32}\mathcal{R}_{\rho\sigma}\mathcal{R}^{\rho\sigma}\right)\gamma_{\mu\nu}^{(0)}
\end{split}
\end{equation}
Note that the authors of \cite{deHaro:2000vlm, Balasubramanian:1999re} show that in $d=6$, $\bar{\gamma}^{(d)}$ is regularization scheme dependent and can be cancelled by a local counterterm.\footnote{The ambiguity comes from a potential $R^2$ term in the counter term action leading to a contribution to the trace of the form $\Box R$. The expression above is for a specific choice of regularization scheme.} Furthermore, this term obeys
\begin{equation}
\gamma^{(0)\mu\nu}\bar{\gamma}^{(d)}_{\mu\nu}=0
\end{equation}
and does not contribute to the conformal anomaly. For the d=6 stress tensor, we will not include this term in our expression. 

The Einstein equations determine the $\gamma^{(i)}$ up to order $z^d$ where new data first appears, including odd $d$. The new data appears in the function $\gamma^{(d)}_{\mu\nu}$ which must be determined from our numerical solution. From the Fefferman-Graham expansion, we can find the boundary stress tensor from the coefficients $\gamma^{(i)}_{\mu\nu}$. For odd d, this is simple to evaluate
\begin{equation}
\langle T^{(d)}_{\mu\nu} \rangle = \frac{dl^{d-1}}{16\pi G_{d+1}}\gamma^{(d)}_{\mu\nu}.
\end{equation}
For even dimensions, however, the expression is more complicated. The important expressions for this work are the d=4 expression,
\begin{equation}
\langle T^{(4)}_{\mu\nu}\rangle = \frac{l^3}{4\pi G_5}\left[\gamma^{(4)}_{\mu\nu}-\frac{1}{8}\left((Tr\gamma^{(2)})^2 - Tr(\gamma^{(2)})^2\right)\gamma^{(0)}_{\mu\nu}-\frac{1}{2}\gamma^{(2)\rho}_{\mu}\gamma_{\nu\rho}^{(2)}+\frac{1}{4}Tr(\gamma^{(2)})\gamma^{(2)}_{\mu\nu}+\frac{3}{2}\bar{\gamma}^{(4)}_{\mu\nu}\right].
\end{equation}
Finally, the six dimensional stress tensor is given by (up to a term proportional to $\bar{\gamma}^{(6)}_{\mu\nu}$. 
\begin{equation}
\label{eq:6dstresstensor}
\langle T^{(6)}_{\mu\nu} \rangle = \frac{3l^5}{8\pi G_7}(\gamma^{(6)}_{\mu\nu} - A^{(6)}_{\mu\nu} + \frac{1}{24} S_{\mu\nu})
\end{equation}
where 
\begin{equation}
\begin{split}
A^{(6)}_{\mu\nu} &= \frac{1}{3}\biggl((\gamma^{(4)}\gamma^{(2)})_{\mu\nu}-(\gamma^{(2)^3}_{\mu\nu}+\frac{1}{8}\left[Tr(\gamma^{(2)^2})-(Tr(\gamma^{(2)}))^2\right]\gamma^{(2)}_{\mu\nu}-Tr(\gamma^{(2)})[\gamma^{(4)}_{\mu\nu}-\frac{1}{2}(\gamma^{(2)^2})_{\mu\nu}]\\
&-\left[\frac{1}{8}Tr(\gamma^{(2)^2})Tr(\gamma^{(2)})-\frac{1}{24}(Tr(\gamma^{(2)}))^3-\frac{1}{6}Tr(\gamma^{(2)^3})+\frac{1}{2}Tr(\gamma^{(2)}\gamma^{(4)})\right]\gamma^{(0)}_{\mu\nu}+2(\gamma^{(2)}\gamma^{(4)})_{\mu\nu}\biggr)
\end{split}
\end{equation}
and
\begin{equation}
\begin{split}
S_{\mu\nu} &= \nabla^2C_{\mu\nu}+2R_{\nu\rho\mu\sigma}C^{\sigma\rho}+4(\gamma^{(2)}\gamma^{(4)}-\gamma^{(4)}\gamma^{(2)})_{\mu\nu}+\frac{1}{10}(\nabla_\mu\nabla_\nu B-\gamma^{(0)}_{\mu\nu}\nabla^2B)\\
&+\frac{2}{4}\gamma^{(2)}_{\mu\nu}B+\gamma^{(0)}_{\mu\nu}(-\frac{2}{3}Tr(\gamma^{(2)^3})-\frac{4}{15}(Tr\gamma^{(2)})^3+\frac{3}{5}Tr\gamma^{(2)}Tr(\gamma^{(2)^2})
\end{split}
\end{equation}
where 
\begin{equation}
C_{\mu\nu} = \left(\gamma^{(4)}-\frac{1}{2}\gamma^{(2)^2}+\frac{1}{4}\gamma^{(2)}Tr(\gamma^{(2)})\right)_{\mu\nu}+\frac{1}{8}\gamma^{(0)}_{\mu\nu}\left( Tr(\gamma^{(2)^2}-(Tr\gamma^{(2)})^2\right).
\end{equation}
In the above expressions, indices are raised with $\gamma^{(0)\mu\nu}$ and lowered with $\gamma^{(0)}_{\mu\nu}$. Expressions like $\gamma^{(2)^2}_{\mu\nu}$ mean $\gamma^{(0)\rho\sigma}\gamma^{(2)}_{\mu\rho}\gamma^{(2)}_{\nu\sigma}$.

To find the coefficients $\gamma_{\mu\nu}^{(i)}$, we need to find an expression for the coordinate $z$ in terms of $x$ and $r$ as well as boundary expansions for $X$ and boundary radial coordinate $R$. To do so, we write
\begin{equation}
\begin{split}
z &= (1-x^2)\left(\frac{1}{1-r^2} + \sum_{n=1}^{\infty} z_{n}(r)(1-x^2)^{n}\right),\\
R&= \frac{R_+}{1-r^2} + \sum_{n=1}^{\infty} R_{n}(r)(1-x^2)^{n},\\
X&=X_0(r) +  \sum_{n=1}^{\infty} X_{n}(r)(1-x^2)^{n} + \log(1-x^2)\sum_{n=1}^{\infty} \tilde{X}_n(r)(1-x^2)^{n},
\end{split}
\end{equation}
where $X_0(r)$ are our Dirichlet boundary conditions (\ref{eq:conformalBCs}). The expansion coefficients $\tilde{X}_n(r)$ are non-zero for $n\leq d$ in $d=4, 6$.\footnote{There are other logarithmic terms that appear at higher order, for instance in $F(x,r)$, where such terms appear at $(1-x^2)^5$ in $d=4$.} We insert the expansion for $X$ into the DeTurck equations and match to the known Fefferman-Graham coefficients $\gamma^{(i)}_{\mu\nu}$. From this, we can find the functions $z_n, R_n$ and $X_n$. Characteristic of asymptotically locally Anti-de Sitter spacetimes, the resulting polynomial contains only even powers of $(1-x^2)$ up to $(1-x^2)^d$. We will omit presenting $z_n, \;R_n,\;B_n$ and $F_n$ because we don't use them explicitly to calculate the stress tensor. 

Because they will be useful for the following, we recall that
\begin{equation}
\begin{split}
\delta_d(r) &= \frac{1}{r^2}(1-(1-r^2)^{d-3})\left(1-(1-r^2)^{d-3}\left(\frac{R_-}{R_+}\right)^{d-3}\right),\\
\delta_4(r) &=(1-(1-r^2)R_-/R_+),\\
\delta_5(r) &=(2-r^2)(1-(1-r^2)^2(R_-/R_+)^2),\\
\delta_6(r) &=(3+3r^2-r^4)(1-(1-r^2)^3(R_-/R_+)^3).
\end{split}
\end{equation}
In $d=4$, the relevant boundary expansion is
\begin{equation}
\begin{split}
\label{eq:asymptotic4}
T&\to \delta_4(r)\left(1-(1-x^2)^2\alpha(r)\right)+(1-x^2)^4 \left(T_4(r)+\log(1-x^2)+\tilde{T}_4(r)\right) + ...,\\
S&\to 1 + \frac{1}{2}(1-x^2)^2\alpha(r)+(1-x^2)^4 \left(S_4(r)+\log(1-x^2)\tilde{S}_4(r)\right) + ... ,\\
A&\to \frac{1}{\delta_4(r)}\left(1 -(1-x^2)^2\alpha(r)\right) + (1-x^2)^4 \left(A_4(r)+\log(1-x^2)\tilde{A}_4(r)\right)+ ...,
\end{split}
\end{equation}
where
\begin{equation}
\alpha(r) = (\delta_4(r)+r^2 R_-/R_+)(1-r^2)
\end{equation}
and the ''$...$" indicates terms of $\mathcal{O}((1-x^2)^5)$ and higher. For completeness, though they don't appear in the stress tensor, the $\tilde{X}_4$ coefficients can be found analytically to be
\begin{equation}
\begin{split}
\tilde{T}_4(r) &= -\frac{3}{2}r^2 \left(r^2-1\right)^2\left(\left(r^2-1\right) R_{-}/R_{+}+1\right)^2R_{-}/R_{+},\\
\tilde{S}_4(r) &=r^2 \left(r^2-1\right)^2 \left(\left(r^2-1\right) R_-/R_++1\right)R_-/R_+,\\
\tilde{A}_4(r) &= -\frac{1}{2} r^2 \left(r^2-1\right)^2 R_-/R_+.
\end{split}
\end{equation}

One can check that as $R_-\to 0$, the above expansion matches Figueras et al.
\newline
\newline
In $d=5$,
\begin{equation}
\begin{split}
\label{eq:asymptotic5}
T&\to \delta_5(r)\left(1 + \frac{3}{4}(1-x^2)^2\beta(r)\right)+(1-x^2)^4\eta_t(r) + (1-x^2)^5 T_5(r) +...,\\
S&\to 1 -\frac{1}{4}(1-x^2)^2\beta(r)+(1-x^2)^4\eta_s(r) + (1-x^2)^5S_5(r) +...,\\
A&\to \frac{1}{\delta_5(r)}\left(1 + \frac{3}{4}(1-x^2)^2\beta(r)\right)+(1-x^2)^4\eta_r(r)+(1-x^2)^5A_5(r) +...,
\end{split}
\end{equation}
where the ''$...$" indicates terms that are $\mathcal{O}((1-x^2)^6)$ and higher. For ease of reading, we have introduced the functions
\begin{equation}
\beta(r) = (1-r^2)^2 (-2 + (3 + 5 r^2 (-2 + r^2)) \frac{R_-^2}{R_+^2})
\end{equation}
and 
\begin{equation}
\begin{split}
\eta_t(r) &= \frac{\delta_5(r)}{112}(-1 + r^2)^2 \\
&\times \Biggl[44 + 204 r^2 (-2 + r^2) - 
   4 (76 + r^2 (-2 + r^2) (589 + 553 r^2 (-2 + r^2))) \frac{R_-^2}{R_+^2}\\
    &\quad\quad\quad\quad+ (-1 + 
      r^2)^2 (227 + r^2 (-2 + r^2) (2258 + 2235 r^2 (-2 + r^2))) {R_-^4}{R_+^4}\Biggr],\\
\eta_s(r) &= \frac{1}{112} (-1 + r^2)^2\\
&\times \Biggl[44 + 40 r^2 - 20 r^4 + 
   4 (-13 + r^2 (-2 + r^2) (132 + 161 r^2 (-2 + r^2)))\frac{R_-^2}{R_+^2}\\
    &\quad\quad\quad\quad- (-1 + 
      r^2)^2 (-59 + r^2 (-2 + r^2) (486 + 565 r^2 (-2 + r^2)))\frac{R_-^4}{R_+^4}\Biggr],\\
\eta_r(r) &=\frac{1}{112\delta_5(r)}(-1 + r^2)^2\\
&\times \Biggl[4 (11 + 37 r^2 (-2 + r^2)) - 
   4 (76 + r^2 (-2 + r^2) (379 + 329 r^2 (-2 + r^2))) \frac{R_-^2}{R_+^2}\\
    &\quad\quad\quad\quad+ (-1 + 
      r^2)^2 (227 + r^2 (-2 + r^2) (1474 + 1395 r^2 (-2 + r^2))) \frac{R_-^4}{R_+^4}\Biggr].\\
\end{split}
\end{equation}
There are no logarithmic terms in $d=5$. 

In d=6, the expansion of X terms are
\begin{equation}
\begin{split}
\label{eq:asymptotic6}
T&\to \delta_6(r)\left(1 + \frac{2}{5}(1-x^2)^2\psi(r)\right)+(1-x^2)^4\chi_t(r) + (1-x^2)^6 \left(T_6(r)+\log(1-x^2)\tilde{T}_6(r)\right)+...,\\
S&\to 1-\frac{1}{10}(1-x^2)^2\psi(r)+(1-x^2)^4\chi_s(r) + (1-x^2)^6\left(S_6(r)+\log(1-x^2)\tilde{S}_6(r)\right)+...,\\
A&\to \frac{1}{\delta_6(r)}\left(1 + \frac{2}{5}(1-x^2)^2\psi(r)\right)+(1-x^2)^4\chi_r(r)+(1-x^2)^6\left(A_6(r)+\log(1-x^2)\tilde{A},_6(r)\right)+...\\
\end{split}
\end{equation}
where the ''$...$" indicates terms that are $\mathcal{O}((1-x^2)^7)$ and higher. For ease of reading, we have defined
\begin{equation}
\psi(r) =  \left(r^2-1\right)^3 \left(\left(14 r^2 \left(r^4-3 r^2+3\right)-9\right) \frac{R_-^3}{R_+^3}+5\right)
\end{equation}
and 
\begin{equation}
\begin{split}
\chi_t(r)&=\frac{\delta_6(r)}{1000}\left(r^2-1\right)^3\\
&\times \biggl[\left(r^2-1\right)^3 \left(\left(r^4-3 r^2+3\right) \left(38374 r^2 \left(r^4-3 r^2+3\right)-40113\right) r^2+5139\right) \frac{R_-^6}{R_+^6}\\&\quad\quad\quad+5 \left(\left(r^4-3 r^2+3\right) \left(7327 r^2 \left(r^4-3 r^2+3\right)-8184\right) r^2+1332\right) \frac{R_-^3}{R_+^3}\\&\quad\quad\quad+25 \left(136 r^2 \left(r^4-3 r^2+3\right)-41\right)\biggr],\\
\chi_s(r)&=\frac{1}{500} \left(r^2-1\right)^3 \\
&\times \biggl[-\left(r^2-1\right)^3 \left(r^2 \left(r^4-3 r^2+3\right) \left(3388 r^2 \left(r^4-3 r^2+3\right)-2781\right)-432\right)  \frac{R_-^6}{R_+^6}\\
&\quad-5 \left(r^2 \left(r^4-3 r^2+3\right) \left(799 r^2 \left(r^4-3 r^2+3\right)-688\right)-36\right)  \frac{R_-^3}{R_+^3}\\
&\quad\quad\quad-25 \left(7 \left(r^4-3 r^2+3\right) r^2+8\right)\biggr],\\
\chi_r(r)&=\frac{1}{1000\delta_6(r)}\left(r^2-1\right)^3 \\
&\times\biggl[\left(r^2-1\right)^3 \left(\left(r^4-3 r^2+3\right) \left(27874 r^2 \left(r^4-3 r^2+3\right)-30363\right) r^2+5139\right) \frac{R_-^6}{R_+^6}\\
&\quad\quad\quad+5 \left(\left(r^4-3 r^2+3\right) \left(5077 r^2 \left(r^4-3 r^2+3\right)-6084\right) r^2+1332\right)\frac{R_-^3}{R_+^3}\\&\quad\quad\quad\quad+25 \left(106 r^2 \left(r^4-3 r^2+3\right)-41\right)\biggr].\\
\end{split}
\end{equation}
The logarithmic terms are
\begin{equation}
\begin{split}
\tilde{T}_6(r) &=\frac{1}{50} f(r)^6 p(r) (R_-/R_+)^3 \left(2 f(r)^9 p(r) \left(9394 r^2 p(r)-1917\right) r^2+63\right) (R_-/R_+)^9\\&\quad+15 f(r)^6 \left(p(r) \left(3502 r^2p(r)-1167\right) r^2+42\right) (R_-/R_+)^6\\
&\quad+6 f(r)^3 \left(p(r) \left(8137 r^2 p(r)-3876\right) r^2+189\right) (R_-/R_+)^3\\
&\quad-5 \left(p(r) \left(3016 r^2 p(r)-1917\right) r^2+126\right),\\
\tilde{S}_6(r) &=\frac{1}{50} \biggl[\left(p(r)\left(5551 r^2 p(r)-1194\right) r^2+63\right) f(r)^{12} (R_-/R_+)^9\\
&\quad+\left(p(r) \left(9908 r^2 p(r)-4305\right) r^2+252\right) f(r)^9 (R_-/R_+)^6\\
&\quad+5 \left(p(r) \left(884 r^2 p(r)-597\right) r^2+63\right) f(r)^6 (R_-/R_+)^3\biggr],
\end{split}
\end{equation}
and
\begin{equation}
\begin{split}
\tilde{A}_6(r) &= \frac{1}{50 p(r)(f(r)^3(R_-/R_+)^3+1)}\biggl(2 \left(p(r)\left(1708 r^2 p(r)-471\right) r^2+63\right) f(r)^{12}(R_-/R_+)^9\\
&\quad-\left(p(r)\left(5890 r^2 p(r)-3549\right) r^2+504\right) f(r)^9 (R_-/R_+)^6\\
&\quad+5 \left(\left(r^4-3 r^2+3\right) p(r) \left(520 r^2 p(r)-471\right) r^2+126\right) f(r)^6 (R_-/R_+)^3\biggr).
\end{split}
\end{equation}

The d=6 stress tensor is
\begin{equation}
\begin{split}
\langle T^{\mu}_{\nu}\rangle &= \frac{3}{8\pi G_7} \text{diag}\biggl\{T^t_{\;\;t},\;\;T^R_{\;\;R},\;\;T^{\Omega}_{\;\;\Omega},\;\;T^{\Omega}_{\;\;\Omega},\;\;T^{\Omega}_{\;\;\Omega},\;\;T^{\Omega}_{\;\;\Omega}\biggr\}\\
\end{split}
\end{equation}
where
\begin{equation}
\begin{split}
(T^t_{\;\;t})R^6&=\frac{\text{T}_6(R)}{\left(1+\frac{R_+}{R}+\frac{R_+^2}{R^2}\right) \left(1-\frac{R_-^3}{R^3}\right)}\\&+\frac{1}{60000 R^{18}}\biggl[166250 R^{15} \left(R_-^3+R_+^3\right)-625 R^{12} \left(827R_-^6+6638 R_-^3R_+^3+827R_+^6\right)\\&+25 R^9 \left(13335 R_-^9+366271 R_-^6R_+^3+366271 R_-^3R_+^6+13335R_+^9\right)-8666482 R_-^9R_+^9\\&-5 R^6R_-^3R_+^3 \left(1011065 R_-^6+4403462 R_-^3R_+^3+1011065R_+^6\right)\\
&+13441280 R^3R_-^6R_+^6 \left(R_-^3+R_+^3\right)\biggr],
\end{split}
\end{equation}
\newline
\begin{equation}
\begin{split}
(T^R_{\;\;R})R^6&=-4 \text{S}_6(R)-\frac{\text{T}_6(R)}{\left(1+\frac{R_+}{R}+\frac{R_+^2}{R^2}\right)\left(1-\frac{R_-^3}{R^3}\right)}\\&+\frac{1}{12000 R^{24}}\biggl[1750 R^{15} \left(R_-^3+R_+^3\right)
+125 R^{12} \left(979 R_-^6-1682 R_-^3R_+^3+979R_+^6\right)\\
&+2863126 R_-^9R_+^9-5 R^9 \left(22575 R_-^9+91367 R_-^6R_+^3+91367R_+^6R_-^3+22575R_+^9\right)\\&+R^6 R_-^3R_+^3 \left(724225 R_-^6+3479278 R_-^3R_+^3+724225R_+^6\right)\\
&-3373640 R^3R_-^6R_+^6 \left(R_-^6+R_+^6\right)\biggr],
\end{split}
\end{equation}
and
\begin{equation}
\begin{split}
(T^{\Omega}_{\;\;\Omega})R^6&=\text{S}_6(R)-\frac{1}{30000 R^{18}}\biggl[21875 R^{15} \left(R_-^3+R_+^3\right)-614681 R_-^9R_+^9
\\&-625 R^{12} \left(19 R_-^6-1040 R_-^3R_+^3+19R_+^6\right)\\
&+175 R^9 \left(165 R_-^9-4909 R_-^6R_+^3-4909 R_-^3R_+^6+165R_+^9\right)\\&+5 R^6 R_-^3R_+^3 \left(35855 R_-^6+134423 R_-^3R_+^3 +35855R_+^6\right)+320365 R^3 R_-^6R_+^6 \left(R_-^3+R_+^3\right)\biggr].
\end{split}
\end{equation}

Finally, the expressions for $\mathcal{A}_d(r)$ are (with $R_+ = 1$)
\begin{equation}
\begin{split}
\mathcal{A}_4(r) &= \frac{1}{4}(\delta_4(r)(-(1-r^2)(R_-r^2+\delta_4(r))(-3+5R_-+5r^2(1+(2r^2-3)R_-))\\
&\quad\quad\quad-8S_4(r))-4T_4(r))/\delta_4(r)^2,\\
 \mathcal{A}_5(r) &=\frac{1}{\delta_5(r)}\left( -3S_5(r) + \frac{T_5(r)}{\delta_5(r)}\right),\\
 \mathcal{A}_6(r) &=\frac{1}{\delta_6(r)}\biggl[-4S_6(r)-\frac{T_6(r)}{\delta_6(r)} \\
 -&\frac{f(r)^3p(r)}{10000}\biggl(-12250R_+^3-17000R_-^3+230580f(r)^3R_-^6-276552f(r)^6R_-^9-75222f(r)^9R_-^12\\
 +&p(r)r^2\bigl(-82375+ 1137150 R_-^3 - 4635720 f(r)^3 R_-^6 + 5992746 f(r)^6 R_-^9 - 2424051 f(r)^9 R_-^{12}\bigr)\\
 +&p(r)^2r^4\bigl(107125 - 2392050R_-^3 + 9669600f(r)^3 R_-^6 - 12945386 f(r)^6 R_-^9 + 5478336 f(r)^9 R_-^{12}\bigr)\\
 +&p(r)^3r^6\bigl(1284400 R_-^3 - 5408835 f(r)^3 R_-^6 + 7468192 f(r)^6 R_-^9 - 3236632  f(r)^9 R_-^{12}\bigr)\biggr)\biggr],
 \label{eq:Ad}
\end{split}
\end{equation}

where $f(r) = 1-r^2$ and $p(r) = 3-3r^2 +r^4$.
\newline

For the sake of completeness, we will also show how to extract the divergences in the entanglement entropies in $d=4$, as an example. The higher dimensional cases are similar. Near the boundary, $z\approx  (1-x^2)R_b$. Next, note that $x^2g(x) = 2x^2-x^4 = 1-z^2/R_b^2$. Finally, $r'(x) = 0$ and $B(x) = 1+\mathcal{O}(z^4)$ near the boundary. Then we can express the area functional near the boundary as
\begin{equation}
\frac{A_4}{4\pi} = R_b^{2}\int_\epsilon \frac{dz}{z}\frac{\left(1+\frac{1}{2}\frac{z^2}{R_b^2}\alpha(r_b)\right)\sqrt{1-z^2/R_b^2}}{z^2} = \frac{R_b^2}{\epsilon^2} - \frac{1}{2}(\alpha(r_b)-1)\text{log}(\epsilon)+\text{finite}.
\end{equation}
It is then just a matter of subtracting the divergent pieces to find the finite entanglement entropy. One must of course check that the finite piece does not vary as a function of the cutoff (for sufficiently small $\epsilon$) which we demonstrate in figure \ref{fig:entanglementerror}.

\bibliography{jammedCFT2}
\bibliographystyle{JHEP}

\end{document}